\documentclass[aps,pra,reprint,groupedaddress,showpacs]{revtex4-1}
\usepackage{graphicx}

\begin{document}

\title{Rotating Fulde-Ferrell-Larkin-Ovchinnikov state in cold Fermi gases}

\author{Tomohiro Yoshida}
\author{Youichi Yanase}
\affiliation{Department of Physics, Niigata University, Niigata 950-2181, Japan}

\date{\today}

\begin{abstract}
We study an effect of rotation on the Fulde-Ferrell-Larkin-Ovchinnikov (FFLO) state
of two component Fermi superfluid gases in a toroidal trap.
We investigate a stability of the FFLO states 
in the quasi-one-dimensional regime on the basis of the Bogoliubov-de 
Gennes equation. 
We find that two novel FFLO phases, i.e., the half quantum vortex state and 
the intermediate state of Fulde-Ferrell (FF) state and Larkin-Ovchinnikov (LO) 
state, are stabilized by the rotation. 
The phase diagram for the FF state, LO state, intermediate state, and
half quantum vortex state is shown in both $T$-$P$ plane and $T$-$h$ plane. 
We demonstrate characteristic features of these states, such as 
the order parameter, flux quantization, and local polarization. 
Several related works are discussed, 
and the advantages of cold Fermi gases are indicated. 
\end{abstract}


\maketitle

\section{INTRODUCTION}
Two component Fermi gases with population imbalance attract much attention
from both theoretical and experimental point of view~\cite{Partridge27012006,Zwierlein27012006,RevModPhys.80.1215}.
One of the motivations of current studies is the realization of
Fulde-Ferrell-Larkin-Ovchinnikov (FFLO) 
state~\cite{PhysRev.135.A550,SovPhysJETP.20.762}, 
which is characterized by a spatial oscillation of the order parameter
arising from the center-of-mass momentum of Cooper pairs.
This state is induced by a mismatch of the Fermi surfaces between two component Fermi particles
produced by a population imbalance.
Since many parameters, such as the population imbalance,  
interaction strength, and trap potential, can be experimentally controlled~\cite{RevModPhys.80.1215},
cold Fermi gases are regarded as a promising candidate for the
FFLO superfluid. 
Recent experiment found an evidence for the FFLO superfluid state in the 
elongated harmonic trap~\cite{Liao}, and attract much attention 
in a variety of fields such as the condensed matter physics
~\cite{Nature.425.51,Bianchi_FFLO,JPSJ.76.051005,Kenzelmann19092008,
PhysRevB.76.134504,PhysRevB.76.054517,
PhysRevLett.102.207004,aperis2010,JPSJ.78.114715} 
and nuclear physics~\cite{RevModPhys.76.263} 
where possible FFLO phases have been discussed. 

The FFLO state has an internal degree of freedom arising from the 
inversion symmetry, and therefore, there are several phases of 
the FFLO state. 
Most of theoretical works on the FFLO state are focused on the two phases. 
One is the Fulde-Ferrell (FF) state which is described by the order
parameter $\Delta({\bf r})\propto{\rm e}^{i{\bf q}\cdot{\bf r}}$, 
and the other is the Larkin-Ovchinnikov (LO) state in which  
$\Delta({\bf r})\propto\cos({\bf q}\cdot{\bf r})$. 
The LO state is regarded as a mixture of two FF states with opposite momentum 
$\Delta({\bf r})\propto{\rm e}^{i{\bf q}\cdot{\bf r}}$ and 
$\Delta({\bf r})\propto{\rm e}^{-i{\bf q}\cdot{\bf r}}$. 
It has been shown that the LO state is stable against the FF state in most cases
~\cite{JPSJ.76.051005}.

The FFLO state in cold Fermi gases has been studied in the literature
~\cite{sheehy:060401,yoshida:063601,PhysRevA.72.025601,PhysRevLett.96.110403,JPSJ.76.104006,PhysRevLett.97.120407,PhysRevA.75.023614,Tezuka-Yanase}.
Some of their works investigated the effect of harmonic trap potential which 
plays an essential role for the spatial structure of the FFLO phase. 
They found that the radial FFLO (R-FFLO) state is stable, in which the order parameter changes its sign
along the radial direction around the edge of the harmonic trap. 
The spatial symmetry remaining in the harmonic trap, however, is not broken in the R-FFLO state,
thus it may be difficult to experimentally observe the spatial 
oscillation of order parameter. 
On the other hand, one of us showed that the angular FFLO (A-FFLO) state 
in which the order parameter changes its sign 
along the angular direction is stable in a toroidal trap~\cite{PhysRevB.80.220510}.
Since the A-FFLO state breaks the rotation symmetry, one expects that 
a spatial modulation characteristic of the FFLO state 
can be detected in experiments. 
Furthermore, A-FFLO state is a novel FFLO state in the sense that the rotation symmetry is spontaneously broken. 
This is in sharp contrast to the superconductors where the rotation symmetry 
is broken by the crystal lattice and therefore the A-FFLO state is not 
stabilized. 
Hence, it is interesting to study the response of A-FFLO state 
to the rotation. It is expected that another novel FFLO state is 
stabilized by the rotation.  
A related work has been given by Kashimura {\it et al}. They showed 
that a spontaneous current appears in a toroidal trap with 
a potential barrier~\cite{PhysRevA.84.013609}, but they didn't 
investigate the effect of rotation. 
In this paper, we investigate the FFLO phases induced by the rotation. 
It is expected that those studies will be tested by the experiments 
since such controllable experiment is an advantage of cold atom gases.  

An effect of rotation on the FFLO state has been studied in the context 
of cold Fermi  
gases~\cite{PhysRevA.81.023601,PhysRevA.80.043610,PhysRevB.82.174514}  
as well as in mesoscopic superconductors where the magnetic field plays an
equivalent role to the rotation~\cite{
PhysRevB.76.184519,PhysRevA.80.043610,PhysRevB.79.174514}. 
It has been shown that the nucleation of vortex in the FFLO
superfluid gives rise to intriguing phenomena. 
In contrast to those works, our study is focused on the 
quasi-one-dimensional superfluid formed on the toroidal trap,  
which is an ideal system to study the effect of gauge field 
on the FFLO state. 
The A-FFLO state is regarded as the LO state along the 
quasi-one-dimensional ring, while order parameter of the vortex state is 
the same as that of the FF state. 
In this work, we study how the A-FFLO state formed on the toroidal trap 
is changed by the nucleation of vortex. 
The same issue has been investigated in the context of mesoscopic
superconducting ring under the tilted magnetic field~\cite{PhysRevB.81.014518}, but 
their study overlooked many important results, as we will show later.
We show that the ``giant vortex state'' corresponding to the FF state is
stabilized in the imbalanced gas near the superfluid critical temperature 
$T_{\rm c}$, and the phase transition to the ``intermediate state'' between the FF and LO states 
occurs with decreasing the temperature. 
Furthermore, the ``half quantum vortex state'' becomes stable in a
certain range of the rotation and population imbalance. 
This phase leads to the half quantized flux of mass and 
the unusual Little-Parks oscillation of $T_{\rm c}$. 
We elucidate superfluid properties of these FFLO states,
such as the superfluid order parameter, local polarization, 
and the flux of mass. 

\section{FORMULATION}
We study rotating two component Fermi gases in a toroidal trap, 
in which atoms are loaded on a quasi-one-dimensional ring~\cite{ryu:260401,Sherlock}. 
We assume that the angular velocity is perpendicular to the plane of ring,
leading to ${\bf \Omega}=\Omega\hat{{\bf z}}$.
Then, the Hamiltonian is approximately described by the 
following one-dimensional attractive Hubbard Hamiltonian: 
\begin{eqnarray}
H&=&-t\sum_{j,\sigma}({\rm e}^{i\Phi}\hat{c}^\dagger_{j+1\sigma}\hat{c}_{j\sigma}+{\rm e}^{-i\Phi}
\hat{c}^\dagger_{j\sigma}\hat{c}_{j+1\sigma}) \nonumber \\
&&-|U|\sum_j\hat{n}_{j\uparrow}\hat{n}_{j\downarrow}-\sum_{j,\sigma}\mu_\sigma \hat{n}_{j\sigma},
\label{eq:eq1}
\end{eqnarray}
where $\sigma=\uparrow, \downarrow$ describe two hyperfine states, 
$\hat{n}_{j\sigma}=\hat{c}^\dagger_{j\sigma}\hat{c}_{j\sigma}$ is the 
number operator, $U$ is the coupling constant of attractive interaction,
and $\mu_\sigma$ is the chemical potential for $\sigma$ particles.
The mass of atoms $M$ is related to $t=1/2M$, and 
the phase $\Phi=R\Omega/2t$ denotes the Peierls phase, 
where $R$ is the radius of ring (see Appendix A). 
We take the unit $\hbar=k_{\rm B}=1$ and $t=1$. 

Although one-dimensional models can be solved with use of 
sophisticated analytic or numerical
methods~\cite{PhysRevB.63.140511,orso:070402,tezuka:110403,hu:070403,liu:043605,feiguin:220508,feiguin:076403,rizzi:245105}, 
we adopt the Bogoliubov-de Gennes (BdG)
equation, because our interests are focused on the quasi-one-dimensional
regime where a three-dimensional long range order is achieved~\cite{parish:250403,zhao:063605,PhysRevA.84.033609}. 
Then, the Hamiltonian is reduced to the mean field Hamiltonian 
\begin{eqnarray}
H_{\rm m}&=&-t\sum_{j,\sigma}({\rm e}^{i\Phi}\hat{c}^\dagger_{j+1\sigma}\hat{c}_{j\sigma}+{\rm e}^{-i\Phi}
\hat{c}^\dagger_{j\sigma}\hat{c}_{j+1\sigma})-\sum_{j,\sigma}\mu_\sigma \hat{n}_{j\sigma} \nonumber \\
&&+\sum_j(\Delta_j \hat{c}^\dagger_{j\uparrow}\hat{c}^\dagger_{j\downarrow}
+\Delta^\ast_j \hat{c}_{j\downarrow}\hat{c}_{j\uparrow})+\sum_j \frac{|\Delta_j|^2}{|U|},
\label{eq:eq2}
\end{eqnarray}
where $\Delta_j\equiv -|U|\langle \hat{c}_{j\downarrow}\hat{c}_{j\uparrow}\rangle$ is the order parameter. 
We focus on the weak coupling BCS region, and then the BdG equation is 
quantitatively appropriate. The Hartree term can be neglected there. 
We don't touch the BCS-BEC crossover region where higher order 
corrections beyond the mean field theory play important
roles~\cite{RevModPhys.80.1215,Chen20051,meisner:023629},  
although it is expected that the FFLO state is robust 
there~\cite{PhysRevB.80.220510,bulgac:215301}. 

The mean field Hamiltonian is diagonalized with use of the
following Bogoliubov transformation: 
\begin{eqnarray}
\hat{c}_{j\uparrow}=\sum_\nu u^\nu _{j\uparrow}\hat{\gamma}_\nu, \   
\hat{c}_{j\downarrow}=\sum_\nu u^{\nu\ast} _{j\downarrow}\hat{\gamma}_\nu^\dagger,
\label{eq:eq3}
\end{eqnarray}
where $\hat{\gamma}_\nu^\dagger$ and $\hat{\gamma}_\nu$ are 
the creation and annihilation operators of quasiparticles, respectively.
The wave function of quasiparticles $(u^\nu_{j\uparrow},u^\nu_{j\downarrow})$ satisfies the BdG equation
\begin{eqnarray}
\sum_l
\left[
\begin{array}{cc}
H_{jl\uparrow} & \Delta_j \delta_{jl} \\
\Delta_j^\ast \delta_{jl} & -H_{jl\downarrow}^\ast \\ 
\end{array}
\right]\left[
\begin{array}{c}
u_{l\uparrow}^\nu \\
u_{l\downarrow}^\nu \\
\end{array}
\right]=E_\nu\left[
\begin{array}{c}
u_{j\uparrow}^\nu \\
u_{j\downarrow}^\nu
\end{array}
\right],
\label{eq:eq4}
\end{eqnarray}
where $H_{jl\sigma}=-t{\rm e}^{i\Phi}\delta_{j-1,l}-t{\rm e}^{-i\Phi}\delta_{j+1,l}-\mu_\sigma\delta_{jl}$.

The self-consistent equation for the order parameter is obtained as,
\begin{eqnarray}
\Delta_j=-|U|\sum_\nu u^\nu_{j\uparrow}u^{\nu\ast}_{j\downarrow}f(E_\nu),
\label{eq:eq5}
\end{eqnarray}
where $f(E_\nu)$ is the Fermi distribution function. 
We can refer to the analytic solution of BdG equation 
in one dimension~\cite{machida1984,PhysRevB.84.024503}, 
however, we numerically solve the self-consistent equation 
(Eqs.~(\ref{eq:eq4}) and (\ref{eq:eq5})) to 
investigate the thermodynamic stability of superfluid phases. 
Several self-consistent solutions corresponding to the metastable 
states are obtained. We determine the stable phase by comparing 
the free energy of those states.  
The free energy is evaluated to be
\begin{eqnarray}
F&=&E - TS \nonumber \\
&=&\sum_\nu E_\nu f(E_\nu)+\sum_j\biggl[\frac{|\Delta_j|^2}{|U|}-\mu_\downarrow
\biggr] \nonumber \\
&&+T\sum_\nu\bigl\{f(-E_\nu)\log[f(-E_\nu)]+
f(E_\nu)\log[f(E_\nu)]\bigr\}. \nonumber \\
&&
\label{eq:eq6}
\end{eqnarray}

We solve BdG equations for a fixed chemical potential $\mu_\sigma$ in
the ground canonical ensemble and calculate the particle number and 
population imbalance which are expressed as $N=N_\uparrow+N_\downarrow$ and 
$P=(N_\uparrow-N_\downarrow)/N$, respectively. 
We obtain the number of $\sigma$ particles $N_\uparrow$ and $N_\downarrow$ as 
\begin{eqnarray}
N_\uparrow&=&\sum_jn_{j\uparrow}=\sum_j\sum_\nu|u^\nu_{j\uparrow}|^2f(E_\nu), \nonumber \\
N_\downarrow&=&\sum_jn_{j\downarrow}=\sum_j\sum_\nu|u^\nu_{j\downarrow}|^2[1-f(E_\nu)], 
\label{eq7}
\end{eqnarray}
respectively.
We define the ``magnetic field'' $h=(\mu_\uparrow-\mu_\downarrow)/2$ 
as the difference of chemical potential between $\uparrow$ and 
$\downarrow$ particles. 

\section{NUMERICAL RESULTS}
We set $|U|/t=1.5$ and assume the chemical potential ($\mu=(\mu_\uparrow+\mu_\downarrow)/2=-0.8$) leading to  
$n_{j\sigma} \equiv \langle \hat{n}_{j\sigma}\rangle \alt 0.4$ so that the lattice Hamiltonian appropriately  
describes the continuum gas. 
The following results are obtained for the number of lattice sites 
$N_{\rm L}=200$ with imposing the periodic boundary condition.  
Note that the periodic boundary condition should be satisfied for 
a gas on the ring. 

In all our results the order parameter $\Delta_j$ is approximately 
described as 
\begin{eqnarray}
\Delta_j=\Delta_+{\rm e}^{iq_+j}+\Delta_-{\rm e}^{-iq_- j}, 
\label{eq:eq8}
\end{eqnarray}
where the center-of-mass momentum of Cooper pairs is 
$q_\pm=2\pi m_\pm/N_{\rm L}$ with $m_\pm$ being an integer 
so as to satisfy the periodic boundary condition. 
The FF state is described by a set of order parameter 
$(\Delta_+, \Delta_-) \propto (1,0)$ or $(\Delta_+, \Delta_-) \propto (0,1)$, 
while $(\Delta_+, \Delta_-) \propto (1,\pm 1)$ with $m_+ = m_- = m$
in the LO state. 

\subsection{Rotating FFLO phases}\label{sec:sec3a}
\begin{figure*}[htbp]
  \begin{tabular}{cc}
    \begin{minipage}{0.5\hsize}
      \includegraphics[width=70mm]{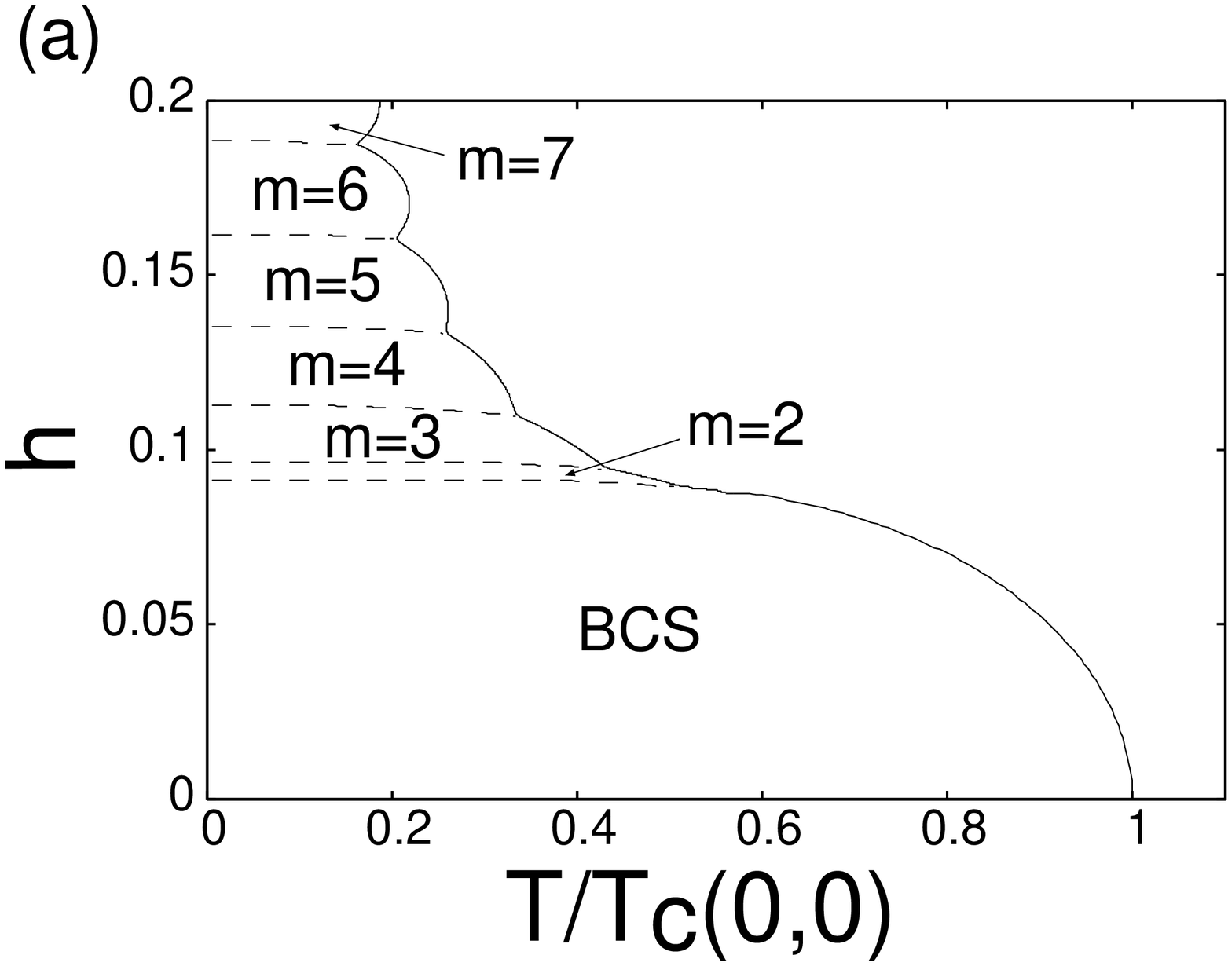}
    \end{minipage}
    \begin{minipage}{0.5\hsize}
      \includegraphics[width=70mm]{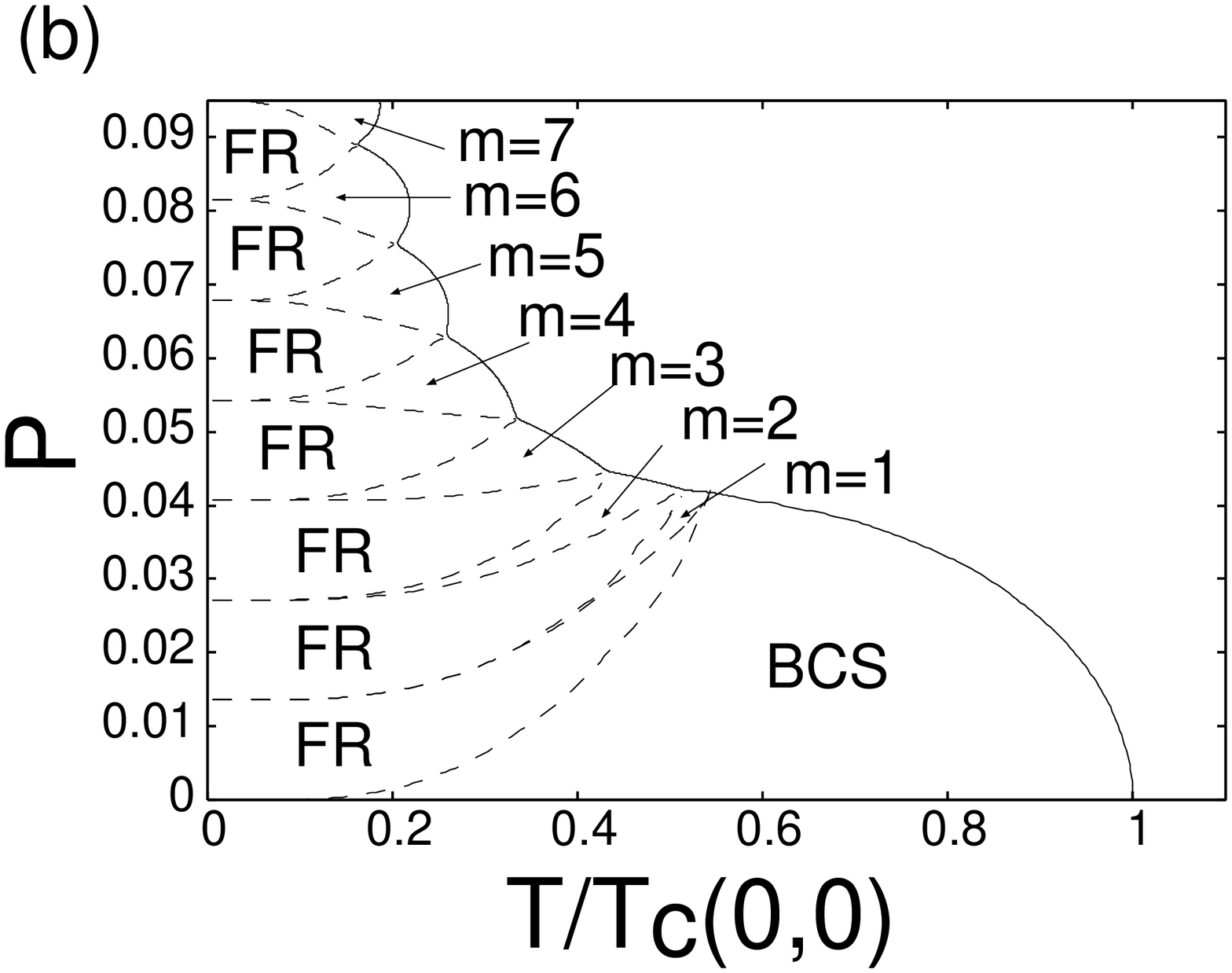}
    \end{minipage}
  \end{tabular}
  \caption{(a) $T-h$ and (b) $T-P$ phase diagram for the gas at rest ($\Phi/\Phi_0=0$).
    The integer $m$ denotes a center-of-mass momentum of Cooper pairs $q=\Phi_0 m$.
    The solid and dashed lines show the second and first order 
    transition lines, respectively. 
    The BCS superfluid state with $m=0$ is denoted as ``BCS'', 
    while $m > 0$ for the LO state is shown in the figure.
    The character ``FR'' denotes the forbidden regimes in Fig.~1(b). 
    The LO state with $m=1$ is stabilized in a tiny region of the $T-h$
    plane, but it cannot be resolved in Fig.~1(a). 
    \label{fig:fig1}}
\end{figure*}

Before studying the effect of rotation, we discuss the phase diagram 
of gases at rest. 
In Fig.~\ref{fig:fig1}, 
we show the phase diagram for $\Phi=0$ 
in the $T-h$ plane as well as in the $T-P$ plane. 
We denote the superfluid critical temperature 
$T_{\rm c}(h,\Phi/\Phi_0)$ as a function of 
the magnetic field $h$ and the normalized phase factor $\Phi/\Phi_0$, 
where $\Phi_0\equiv 2\pi/N_{\rm L}$. 
The temperature in Fig.~1 is normalized by the critical temperature of the balanced 
gas at rest, $T_{\rm c}(0,0) = 0.0816$.  
Figure~\ref{fig:fig1}(a) shows that the BCS state is stable for $h<0.0915$ while the LO state 
is stabilized for $h>0.0915$ at $T=0$. 
Note that the order parameter of the LO state is described as 
$\Delta_j\propto\cos[(\Phi_0 m)j]$. 
Since we consider the Fermi gas trapped on the ring, 
this LO state corresponds to the A-FFLO state studied in 
Ref.~\cite{PhysRevB.80.220510}. 
The integer $m$ increases with increasing the magnetic field 
$h$~\cite{JPSJ.76.051005}. 
Although the FFLO state is stable for a huge magnetic field 
$h \gg T_{\rm c}(0,0)$ in our calculation, 
our study focuses on the moderate magnetic field $h \sim T_{\rm c}$ 
where $m \ll N_{\rm L}$. 
Figure 1(b) shows that the BCS state is unstable in the imbalanced gas $P > 0$ at 
$T=0$. This is because the BCS state does not have an excess 
particle owing to the excitation gap.
We also find forbidden regimes (FR) in the $T-P$ plane since the particle 
number is finite and not fixed in our calculation based on the grand canonical ensemble.  
Another inhomogeneous superfluid state may be stabilized in the canonical ensemble, 
but that is beyond the scope of our paper. 

\begin{figure*}[htbp]
  \begin{tabular}{cc}
    \begin{minipage}{0.5\hsize}
      \includegraphics[width=50mm,angle=270]{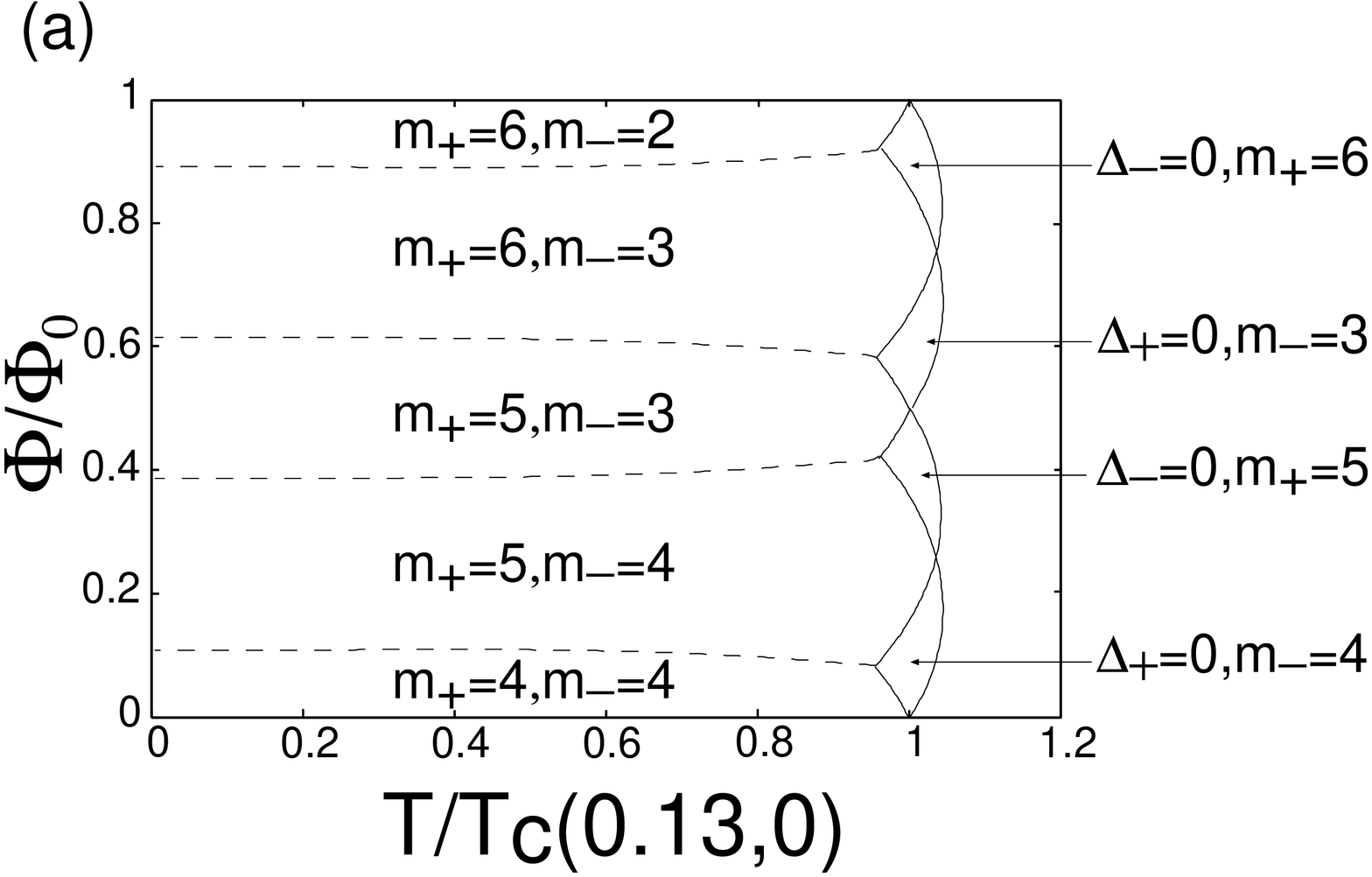}%
    \end{minipage}
    \begin{minipage}{0.5\hsize}
      \includegraphics[width=50mm,angle=270]{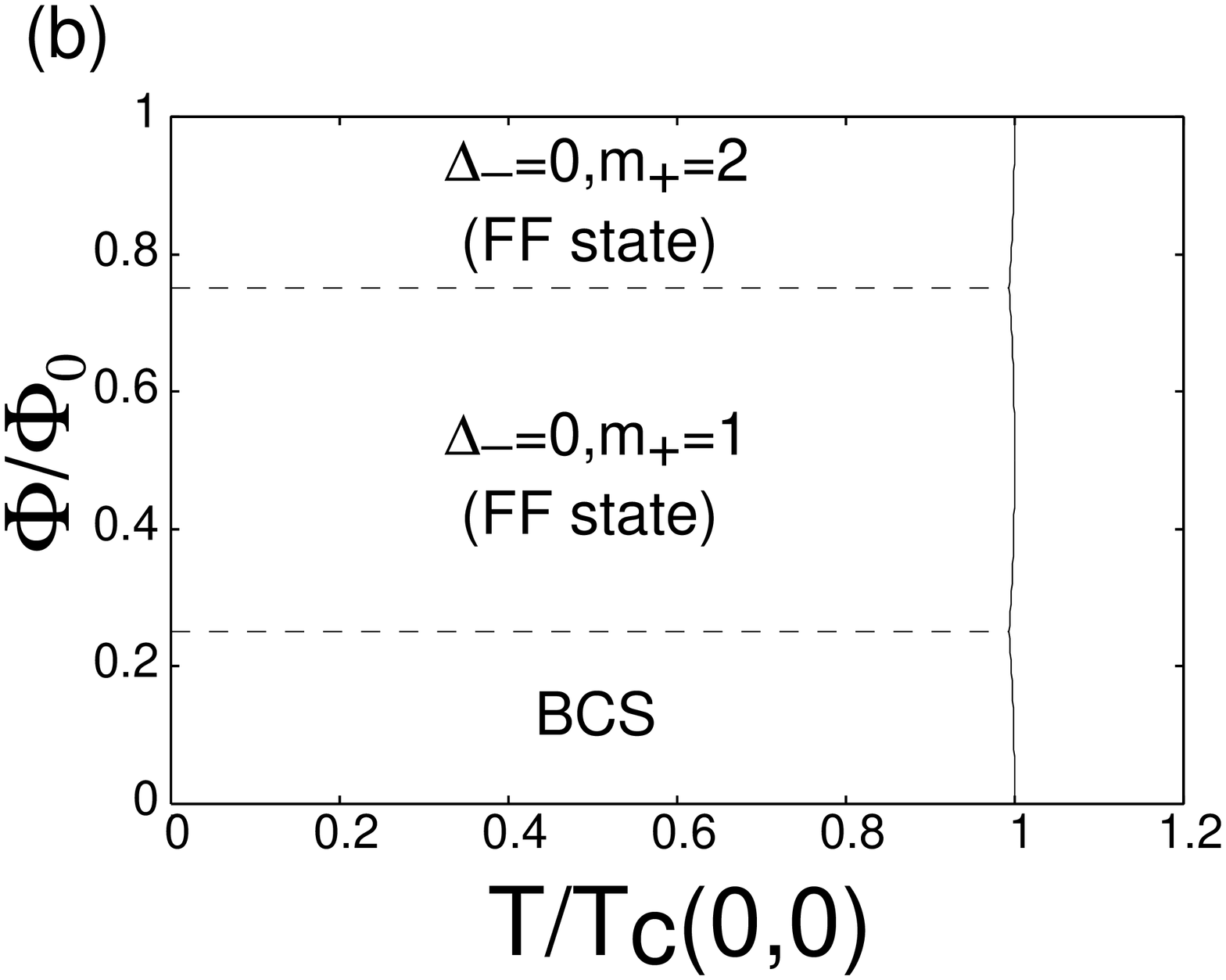}%
    \end{minipage}
  \end{tabular}
    \caption{$T-\Phi$ phase diagram at (a) $h=0.13$ and (b) $h=0$. 
      The solid and dashed lines show the second and first order 
      transition lines, respectively.
      The integers $m_\pm$ denote the momentum of Cooper pairs, and 
      $\Delta_\pm$ are the amplitude of order parameter for
      $m_\pm$, respectively 
      (see Eq.~(\ref{eq:eq8})).
      The phase diagram is periodic for $\Phi$ with the period $ \Phi_{0} $ (see Appendix B). 
      \label{fig:fig2}}
\end{figure*}

We here turn to the rotating systems. 
Figure~\ref{fig:fig2}(a) 
shows the phase diagram in the $T-\Phi$ plane for a field $h=0.13$ 
where the LO state with $m=4$ is stable at rest, 
while Fig.~\ref{fig:fig2}(b) shows the phase diagram 
for $h=0$.
While a conventional vortex state and usual Little Parks oscillation 
of $T_{\rm c}$ appear in the rotating BCS state (Fig.~\ref{fig:fig2}(b)), 
we see several intriguing features in the rotating FFLO state 
(Fig.~\ref{fig:fig2}(a)). 
First, the FF state with $\Delta_+=0$ or $\Delta_-=0$ is stabilized 
near the critical temperature in the rotating gas with $\Phi/\Phi_{0} \neq 0$. 
The two FF states $\Delta_j\propto {\rm e}^{iq j}$ and 
$\Delta_j\propto {\rm e}^{-iq j}$ are degenerate at rest, and therefore, 
the mixture of them, i.e., the LO state 
$\Delta_j\propto ({\rm e}^{iq j} + {\rm e}^{-iq j})$ is stabilized at 
$T=T_{\rm c}$. 
On the other hand, the degeneracy is lifted by the rotation, and then 
one of the FF states is stabilized near $T_{\rm c}$. 
This FF state is regarded as a ``giant vortex state'' 
because the vorticity $m$ is much larger than the conventional one. 
The critical temperature is increased by the rotation since one of the 
FF states is stabilized. 
The FF state may be furthermore stabilized by the Hartree term, which is 
neglected here, because of the Fermi liquid correction~\cite{PhysRevB.74.172504}.

Second, the ``half quantum vortex state'' is stabilized by the rotation. 
Figure~\ref{fig:fig2}(a) shows that the LO state with 
$(m_+, m_-) =(4,4)$ changes to the phase $(m_+, m_-) =(5,4)$, 
as the rotation, i.e., the phase $\Phi$ increases. 
The latter phase is regarded as the half quantum vortex state 
since the order parameter is approximately described at low temperatures as  
\begin{eqnarray}
\Delta_j&\propto& {\rm exp}[i(\Phi_0\times 5)j]+
{\rm exp}[-i(\Phi_0\times 4)j]
 \nonumber \\
&\propto&  {\rm exp}[i(\Phi_0\times 0.5 )j]
\cos[(\Phi_0\times 4.5)j].
\label{eq:9}
\end{eqnarray} 
The exponential provides the phase $\pi$ for a circuit along the ring, 
which corresponds to a half quantized vortex. 
Another phase $\pi$ required for the uniqueness of order parameter 
is given by the odd number of phase changes in the sinusoidal function. 
Then, the order parameter of the half quantum vortex state is regarded 
as the product of those in the FF and LO states with an unconventional period. 
Similarly, the phase with $(m_+, m_-) =(6,3)$ is the half quantum vortex 
state while the phases with $(m_+, m_-) =(5,3)$ and $(m_+, m_-) =(6,2)$ 
are the LO state having one and two quantized vortex, respectively. 
Thus, the half quantum vortex state is realized when the difference of 
vorticity is odd between the coexisting FF states. 
One of the two quantum numbers $m_+$ and $m_-$ changes with the increase 
of rotation, and induces the half quantized vortex. 
This is in sharp contrast to the rotating BCS state which has only 
one quantum number $m$. 
In other words, the half quantized vortex is induced by the multi-component 
order parameters of the FFLO state, as in the chiral $p$-wave 
superconductor~\cite{ivanov2001}. 

\begin{figure}[htbp]
  \includegraphics[width=70mm]{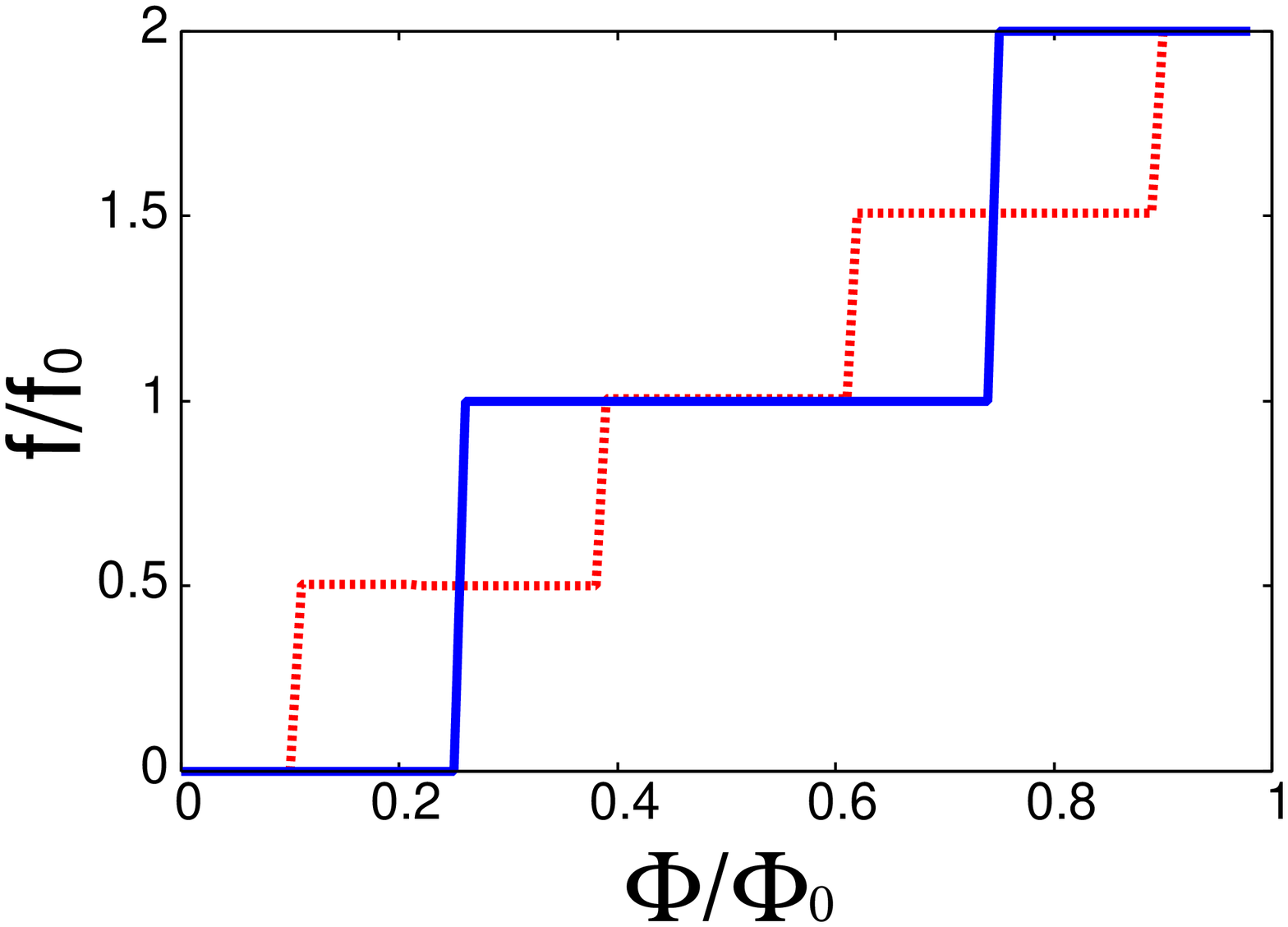}
  \caption{(Color online) The rotation dependence of 
    the flux of mass at $T=0$ in the BCS state ($h=0$, blue solid line) 
    and in the FFLO state ($h=0.13$, red dotted line).
    The flux of mass is normalized by the value at $h=0$ and $\Phi/\Phi_0=0.5$.
    \label{fig:fig3}}
\end{figure}

We see a characteristic feature of the half quantum vortex state  
appearing in the quantized flux of mass. 
We show the rotation dependence of the flux of mass $f$ for 
$h=0$ and $h=0.13$ at $T=0$ in Fig.~\ref{fig:fig3},
which is obtained as
\begin{eqnarray} 
f&=&f_{\uparrow}+f_{\downarrow},\label{eq:eq10} \\
f_{\uparrow}&=&2t\Im\biggl[\sum_\nu u_{j\uparrow}^{\nu\ast}u_{j+1\uparrow}^\nu f(E_\nu)\biggr], 
\label{eq:eq11}
\\
f_{\downarrow}&=&2t\Im\biggl\{\sum_\nu u_{j\downarrow}^\nu u_{j+1\downarrow}^{\nu\ast} [1-f(E_\nu)]\biggr\}.
\label{eq:eq12}
\end{eqnarray}
Since the flux is independent of the site, we omit the index $j$
in Eqs.~(\ref{eq:eq11}) and (\ref{eq:eq12}). 
Figure~\ref{fig:fig3} clearly shows that the flux of mass is 
half quantized in the rotating FFLO state ($h=0.13$). 

Owing to the nucleation of half quantum vortex, 
the $T_{\rm c}$ shows an unconventional Little-Parks 
oscillation studied in Ref.~\cite{PhysRevB.79.174514}. 
Figure~\ref{fig:fig2}(a) 
shows the Little Parks oscillation with the period being 
a half of the conventional one. 
It should be noticed that the amplitude of oscillation 
is remarkably enhanced in the FFLO state compared with the BCS state. 

\begin{figure}[htbp]
  \includegraphics[width=50mm,angle=270]{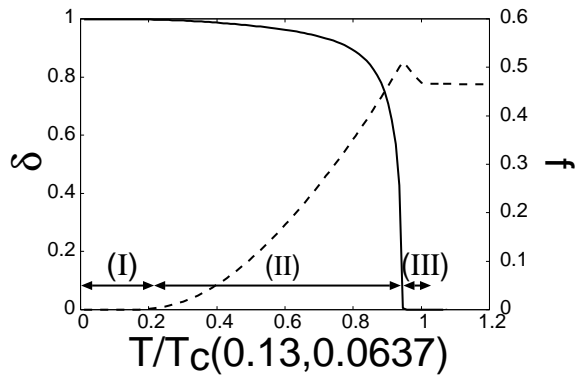}
  \caption{Temperature dependence of the ratio $\delta\equiv \Delta_+/\Delta_-$ (solid line) and
    the flux of mass $f$ (dashed line) at $h=0.13$
    and $\Phi/\Phi_0=0.0637$.
    \label{fig:fig4}}
\end{figure}

Third, we point out another phase in Fig.~\ref{fig:fig2}(a), 
that is the mixture of LO state and FF state. 
For instance, the FFLO state with $(m_+, m_-) = (4,4)$ is not a pure LO state
in the sense that the amplitude $\Delta_+$ for $m_+=4$ 
is not equivalent to $\Delta_-$ for $m_-=4$. 
The temperature dependence of the ratio $\delta\equiv \Delta_+/\Delta_-$ 
at $h=0.13$ and $\Phi/\Phi_0=0.0637$ is shown in 
Fig.~\ref{fig:fig4}, 
where the superfluid phase is divided into three regions, 
(I), (II), and (III). 
In the low temperature region (I), 
the ratio $\delta$ is almost unity corresponding to the LO state, while 
the FF state with $\delta = 0$ is stabilized in the 
high temperature region (III).
Interestingly, the intermediate ratio $0<\delta<1$ 
is obtained in a wide intermediate temperature region (II). 
Thus, the ``intermediate state'' between the FF and LO states is 
stabilized by the rotation, 
and then the order parameter is described as 
\begin{eqnarray}
\Delta_j\propto\delta{\rm exp}[i(\Phi_0\times 4)j]
+{\rm exp}[-i(\Phi_0\times 4)j]. 
\label{eq:eq13}
\end{eqnarray}

Although the intermediate state changes to the LO state through the 
crossover, these states can be distinguished by observing the flux of mass. 
As shown in Fig.~\ref{fig:fig4},  
the flux of mass is almost zero in the LO state. 
This is because the susceptibility to the rotation is suppressed by the 
excitation gap. A narrow band is formed by the Andreev bound states 
localized around the nodes of order parameter~\cite{PhysRevA.84.033609}, 
its contribution is negligible for these parameters. 
On the other hand, the flux of mass increases with temperature  
in the region (II), 
and shows a peak at the transition temperature 
to the FF state, $T = T_{\rm c2} = 0.946 T_{\rm c}(0.13,0.0637)$. 
The flux decreases as approaching to 
the superfluid critical temperature $T=T_{\rm c}$ because 
the order parameter of FF state is decreased. 
In other words, the flux of mass is first enhanced below $T_{\rm c}$ owing to the 
formation of the giant vortex state, and then suppressed below $T_{\rm c2}$ as 
decreasing the temperature. 
Such nonmonotonic temperature dependence is a characteristic feature 
of the FFLO state, while that is not observed in the BCS superfluid. 
Note that we here assume the system without conserving the angular 
momentum. Another phase diagram will be obtained when the 
angular momentum is conserved across the superfluid transition.  

\subsection{Local polarization}

\begin{figure*}[htbp]
  \begin{tabular}{cc}
    \begin{minipage}{0.5\hsize}
      \includegraphics[width=70mm]{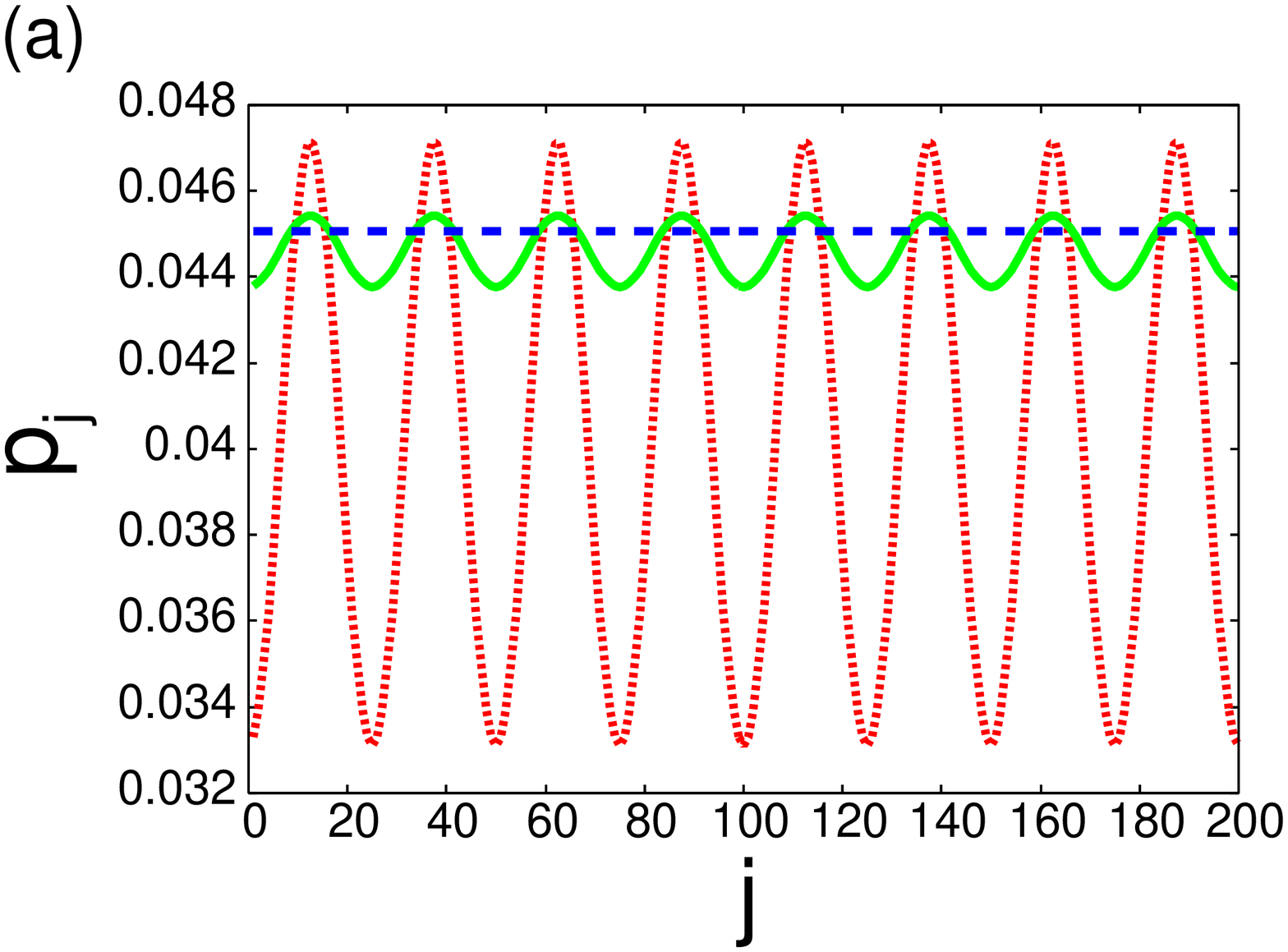}
    \end{minipage}
    \begin{minipage}{0.5\hsize}
      \includegraphics[width=70mm]{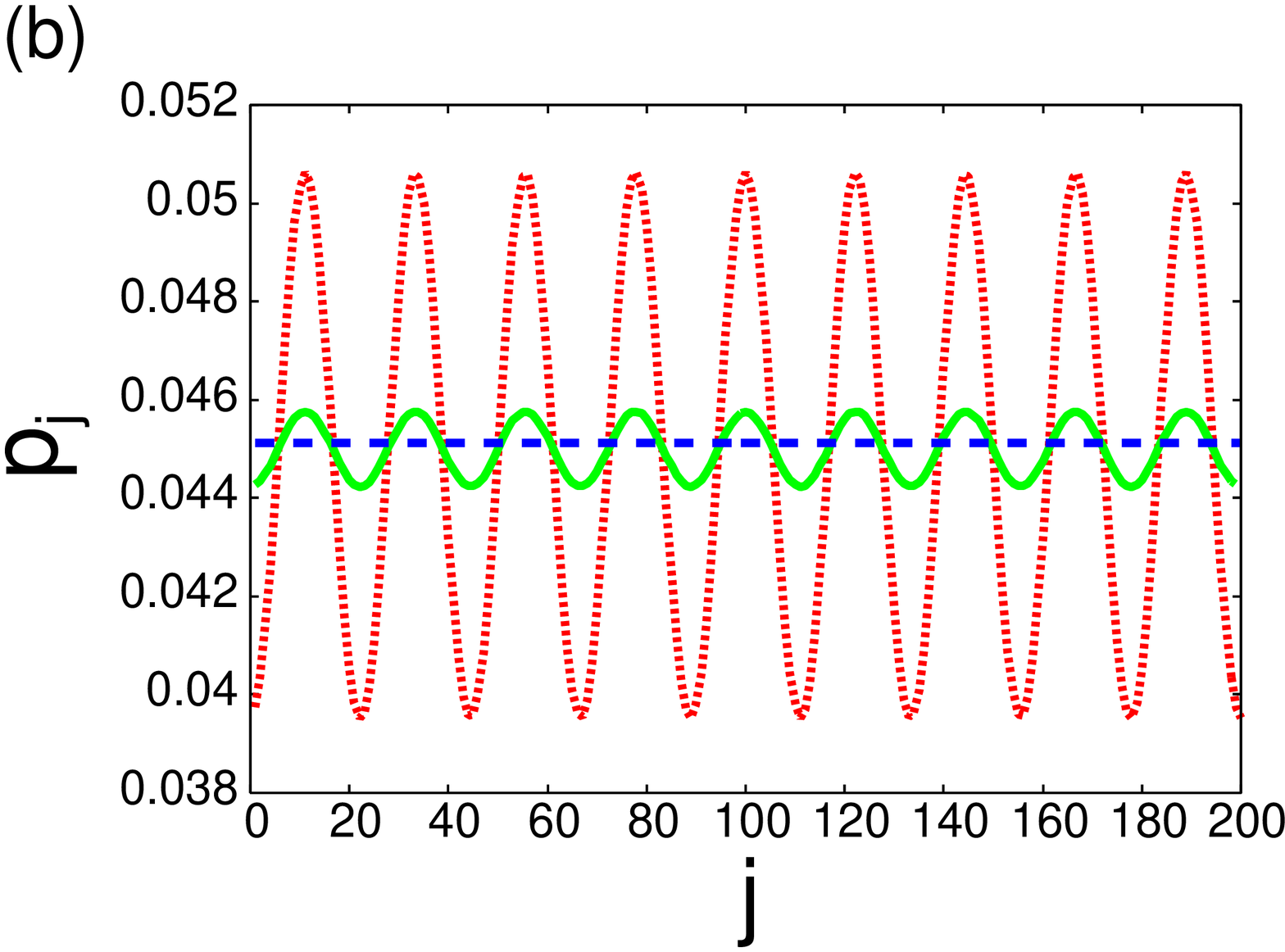}
    \end{minipage}
  \end{tabular}
  \caption{(Color online) Spatial profile of the local polarization
    at $h=0.13$.
    (a) $\Phi/\Phi_0=0.0637$. The dotted (red), 
    solid (green), and dashed (blue) lines represent the data for 
    $T/T_{\rm c}(0.13,0)=0$ (LO state), $T/T_{\rm c}(0.13,0)=0.947$ (intermediate state), and 
    $T/T_{\rm c}(0.13,0)=0.991$ (FF state), respectively. 
    (b) $\Phi/\Phi_0=0.159$. The dotted (red), solid (green), and
    dashed (blue) lines represent the data for 
    $T/T_{\rm c}(0.13,0)=0$ (half quantum vortex state), 
    $T/T_{\rm c}(0.13,0)=0.969$ (intermediate state), and 
    $T/T_{\rm c}(0.13,0)=1.01$ (FF state), respectively.
    \label{fig:fig5}}
\end{figure*}

\begin{figure*}[htbp]
  \begin{tabular}{cc}
    \begin{minipage}{0.5\hsize}
      \includegraphics[width=70mm]{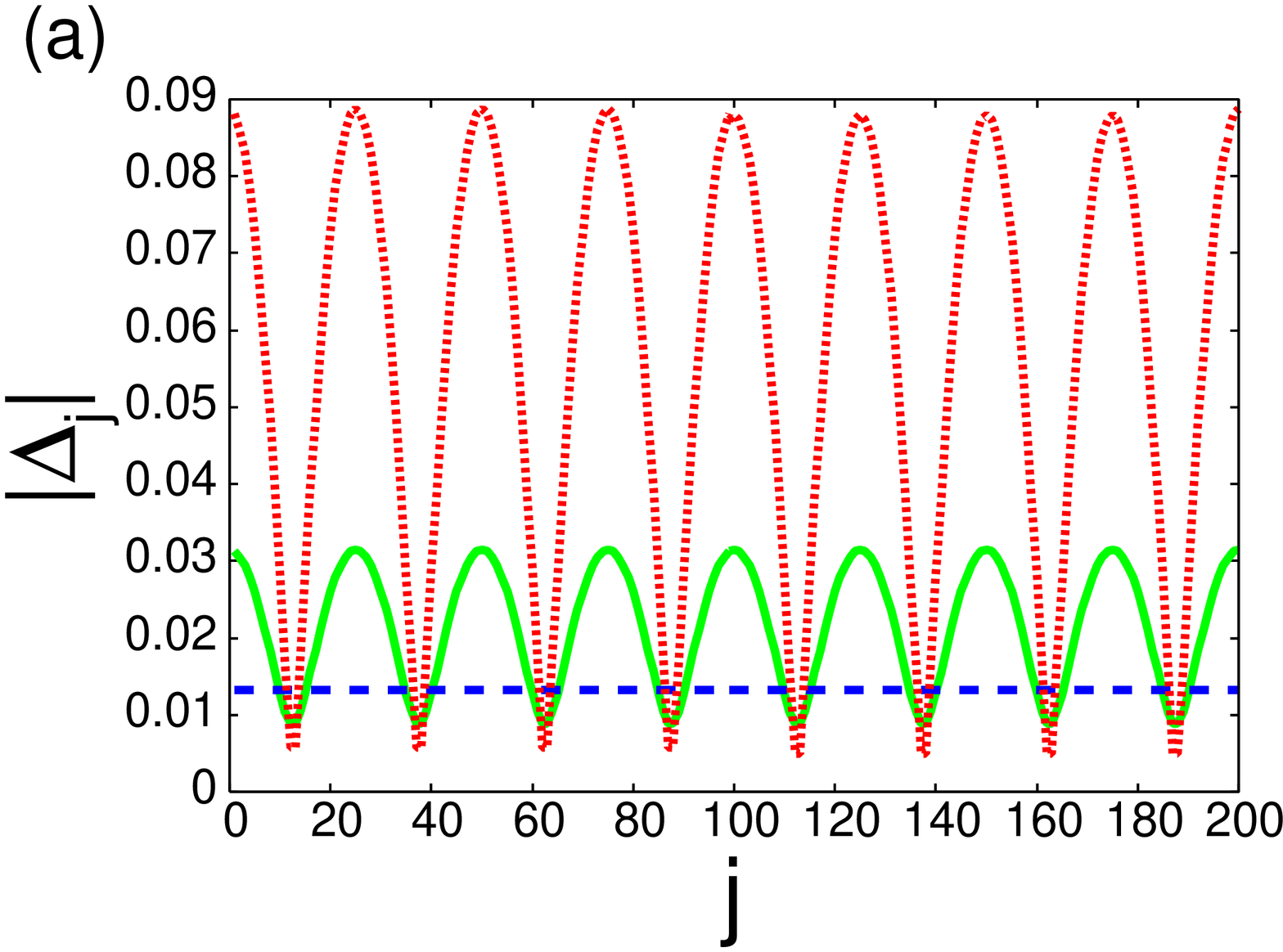}
    \end{minipage}
    \begin{minipage}{0.5\hsize}
      \includegraphics[width=70mm]{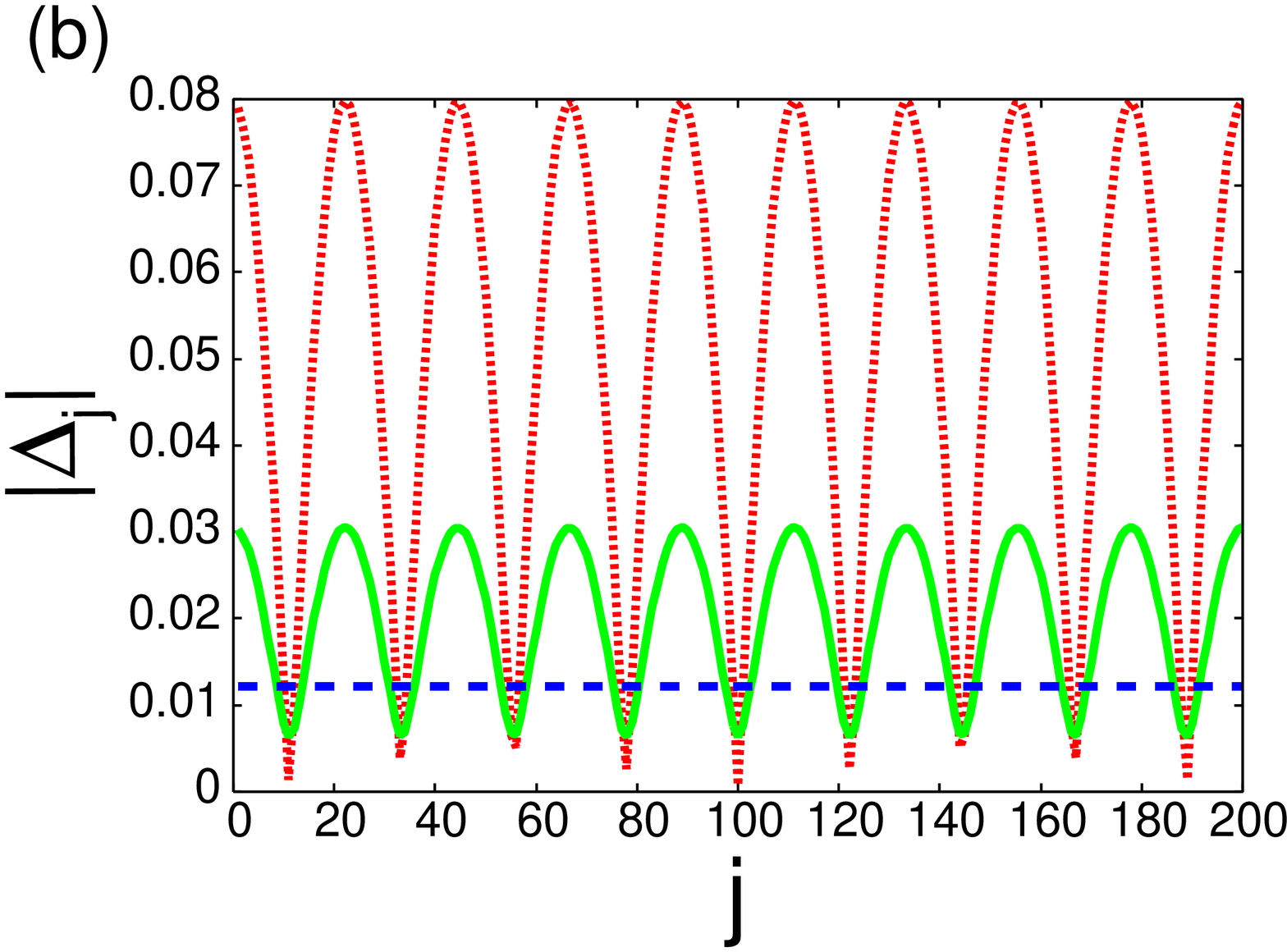}
    \end{minipage}
  \end{tabular}

  \caption{(Color online) Spatial profile of the amplitude 
    of order parameter $|\Delta_j|$  at $h=0.13$.
    In (a) and (b), the parameters are the same as 
    in Figs.~\ref{fig:fig5}(a) and (b), respectively.  
    \label{fig:fig6}
}
\end{figure*}

The FFLO phases studied in the previous subsection 
show their characteristic feature in the local polarization. 
Figure~\ref{fig:fig5} shows the spatial profile of  
local polarization $p_j=n_{j\uparrow}-n_{j\downarrow}$ 
in the FF, LO, half quantum vortex, and intermediate states. 
It is shown that the local polarization in the LO state shows peaks 
at the quasi-nodal points of order parameter where $|\Delta_j|\sim 0$ (compare Fig.~\ref{fig:fig5}
with Fig.~\ref{fig:fig6}), as in the A-FFLO 
state~\cite{PhysRevB.80.220510}.
This is because excess particles are localized around the quasi-nodal points
of order parameter in order to make the loss of condensation energy as small as possible.
Therefore the number of peaks in the local polarization $n_0$ can be regarded as the
number of quasi-nodal points in the order parameter.
On the other hand, the FF state shows a uniform polarization,
since the translation symmetry is not broken there. 
The spatial inhomogeneity is smoothly enhanced by decreasing the temperature 
in the intermediate state. 
Thus, we can discriminate the LO state, intermediate state, and FF state by
the polarization measurements. 

A characteristic feature of the half quantum vortex state 
appears in the number of peaks in the local polarization. 
The number $n_0$ is odd (even) in the half (integer) quantum vortex state.  
(See above discussion and Fig.~\ref{fig:fig6}.) 
For example, Fig.~\ref{fig:fig5}(a) shows $n_0 = 8$, 
while that is $n_0 = 9$ in Fig.~\ref{fig:fig5}(b) where 
the amplitude is described by
$|\Delta_j|\propto |\cos[(\Phi_0\times 4.5)j]|$. 
Thus, the odd number of $n_0$ is a clear signature of the 
half quantum vortex state. 
Note that the $n_0$ is the same as the number of excess particles. 
As increasing the excess particles, the number of quasi-nodes increases 
so as to hold them. 

\subsection{Mesoscopic effect}\label{sec:sec3c}

\begin{figure}[htbp]
  \includegraphics[width=50mm,angle=270]{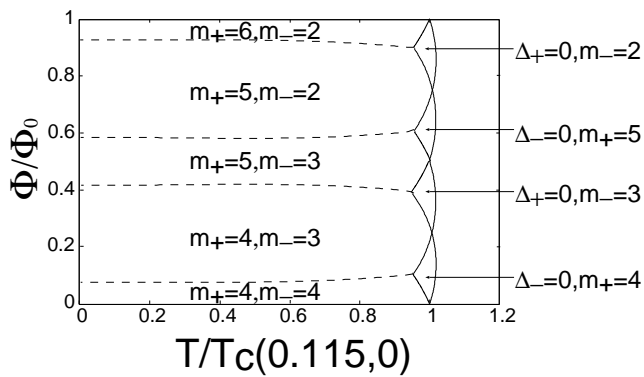}
  \caption{$T-\Phi$ phase diagram at $h=0.115$.
           The other parameters are the same as Fig.~\ref{fig:fig2}.
\label{fig:fig7}}
\end{figure}

We here comment on the magnetic field dependence of superfluid phases. 
Although the LO state with $(m_+, m_-) = (4,4)$ is stabilized at
rest in both magnetic fields $h=0.115$ and $h=0.13$, the FF state 
induced by the rotation depends on the field. 
While the FF state with $\Delta_+ = 0$ is stabilized for $h=0.13$ 
(Fig.~\ref{fig:fig2}(a)), the other FF state with 
$\Delta_- = 0$ is stabilized for $h=0.115$ as shown in 
Fig.~\ref{fig:fig7}. 
Accompanied by the change of FF state, the half quantum vortex state 
also changes as increasing the magnetic field from $h=0.115$ to 
$h=0.13$; a small rotation stabilizes the half quantum vortex state 
$(m_+, m_-) = (4,3)$ in the former, while $(m_+, m_-) = (5,4)$ 
in the latter. 

We understand these magnetic field dependences in the following two 
ways. First, the superfluid phase having a large number of zeros of 
order parameter $n_0$ is stabilized by increasing the magnetic field 
so as to gain the magnetic energy. 
For example, $n_0 = 7$ [$(m_+, m_-) = (4,3)$] at $h=0.115$ 
increases to $n_0 = 9$ [$(m_+, m_-) = (5,4)$] at $h=0.13$. 

Another interpretation is based on the discrete single particle
spectrum, that is the mesoscopic effect.  
\begin{figure*}[htbp]
  \begin{tabular}{cc}
    \begin{minipage}{0.5\hsize}
      \includegraphics[width=70mm]{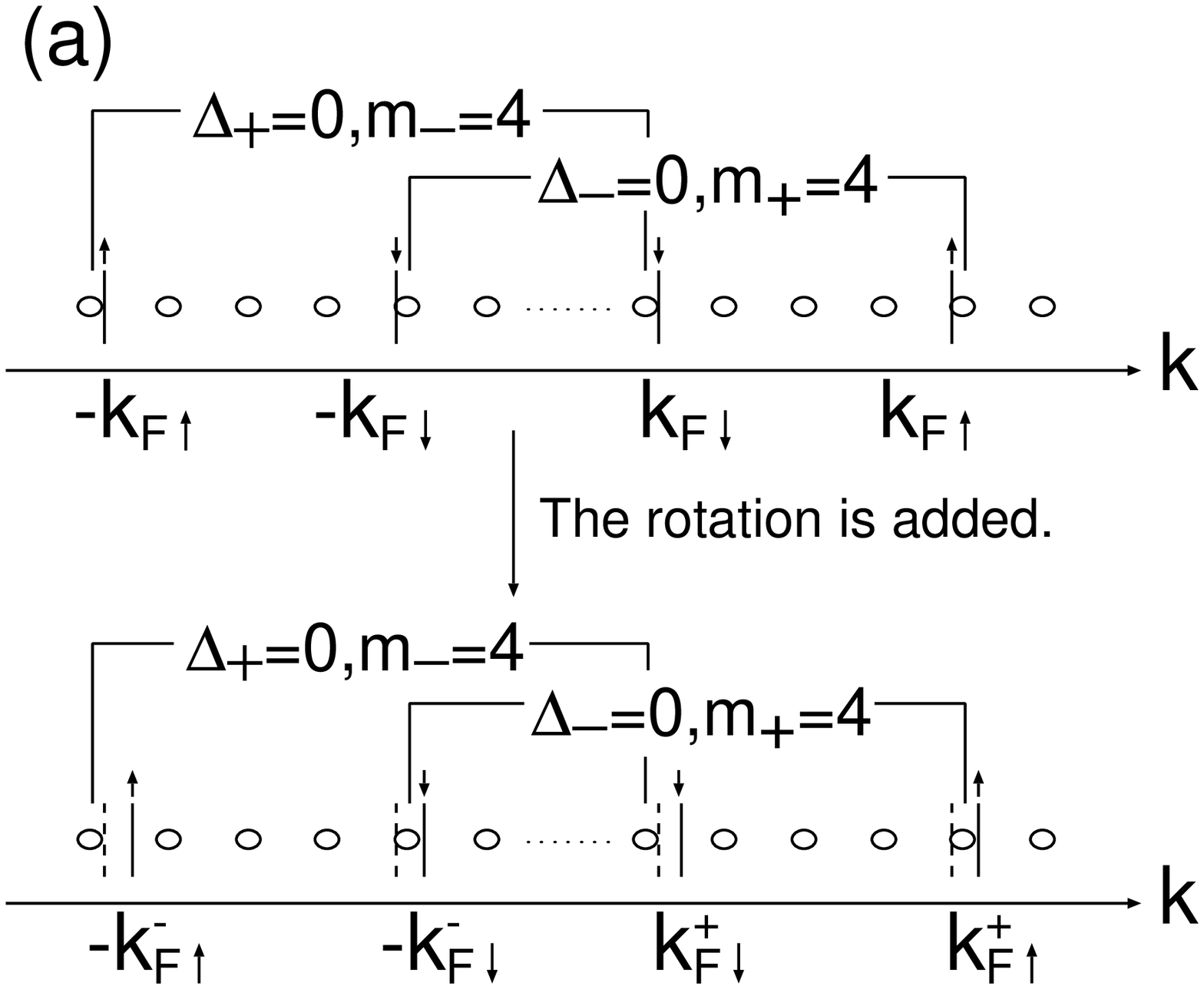}
    \end{minipage}
    \begin{minipage}{0.5\hsize}
      \includegraphics[width=70mm]{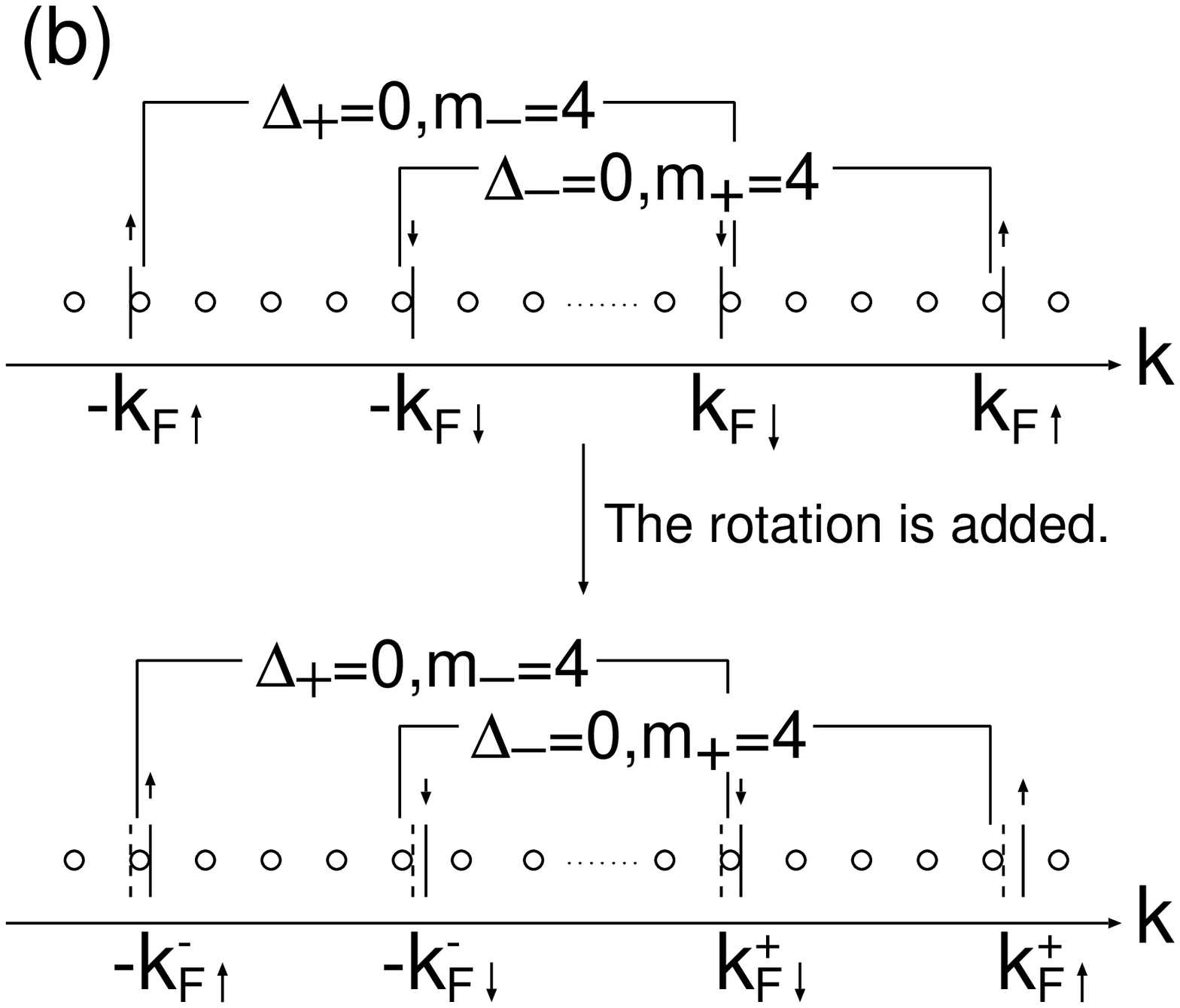}
    \end{minipage}
  \end{tabular}
  \caption{Schematic sketch of Cooper pairing in the FFLO state. 
    (a) the case of $h=0.115$ and (b) the case of $h=0.13$.
    The upper panel shows the quasiparticle energy of the gases at rest,
    while the lower panel shows the shift of Fermi momentum due to the rotation. 
    The circles represent the momentum discretized by the mesoscopic effect. 
    $k_{{\rm F}\sigma}$ and $k_{{\rm F}\sigma}^\pm$ are the Fermi momentum for $\sigma$ particles. 
    \label{fig:fig8}}
\end{figure*}
Figure~\ref{fig:fig8} shows the schematic sketch of 
the Cooper pairing for the cases $h=0.115$ and $h=0.13$. 
The upper panel shows the Fermi momentum at rest, while the lower panel 
shows the shift of Fermi momentum due to the rotation (see Appendix A). 
This shift gives rise to the lifting of 
degeneracy of two FF states with $\Delta_+ =0$ and $\Delta_- =0$. 
In Fig.~\ref{fig:fig8}(a), a small shift moves the Cooper pairs with $m_-=4$ away from the Fermi surface.
Since the Cooper pairs should be formed by the quasiparticles 
near the Fermi surface, 
the FF state with $m_+=4$ is favored at low magnetic fields 
as shown in Fig.~\ref{fig:fig8}(a), while 
the other FF state with $m_-=4$ is stabilized at high magnetic 
fields as shown in Fig.~\ref{fig:fig8}(b). 
In other words, the reconstruction of quasiparticle energy due to the rotation 
determines the stable FF state near $T_{\rm c}$. 
When the temperature is decreased, the sub-dominant order parameter 
appears below $T_{\rm c2}$. 
That is $\Delta_+$ with $m_+=4$ ($\Delta_-$ with $m_- =4$) for a small rotation, 
and changes to $m_+=5$ ($m_- =3$) as increasing the rotation. 
Thus, the FF state with $\Delta_+ =0$ changes 
to the half quantized vortex state $(m_+, m_-) = (5,4)$, 
while the other FF state with $\Delta_- =0$ changes to 
another half quantum vortex state $(m_+, m_-) = (4,3)$. 
Note again that these differences of the $T-\Phi$ phase diagram between 
$h=0.115$ and $h=0.13$ are induced by the mesoscopic effect. 

\subsection{Phase diagram}

\begin{figure*}[htbp]
  \begin{tabular}{cc}
    \begin{minipage}{0.5\hsize}
      \includegraphics[width=70mm]{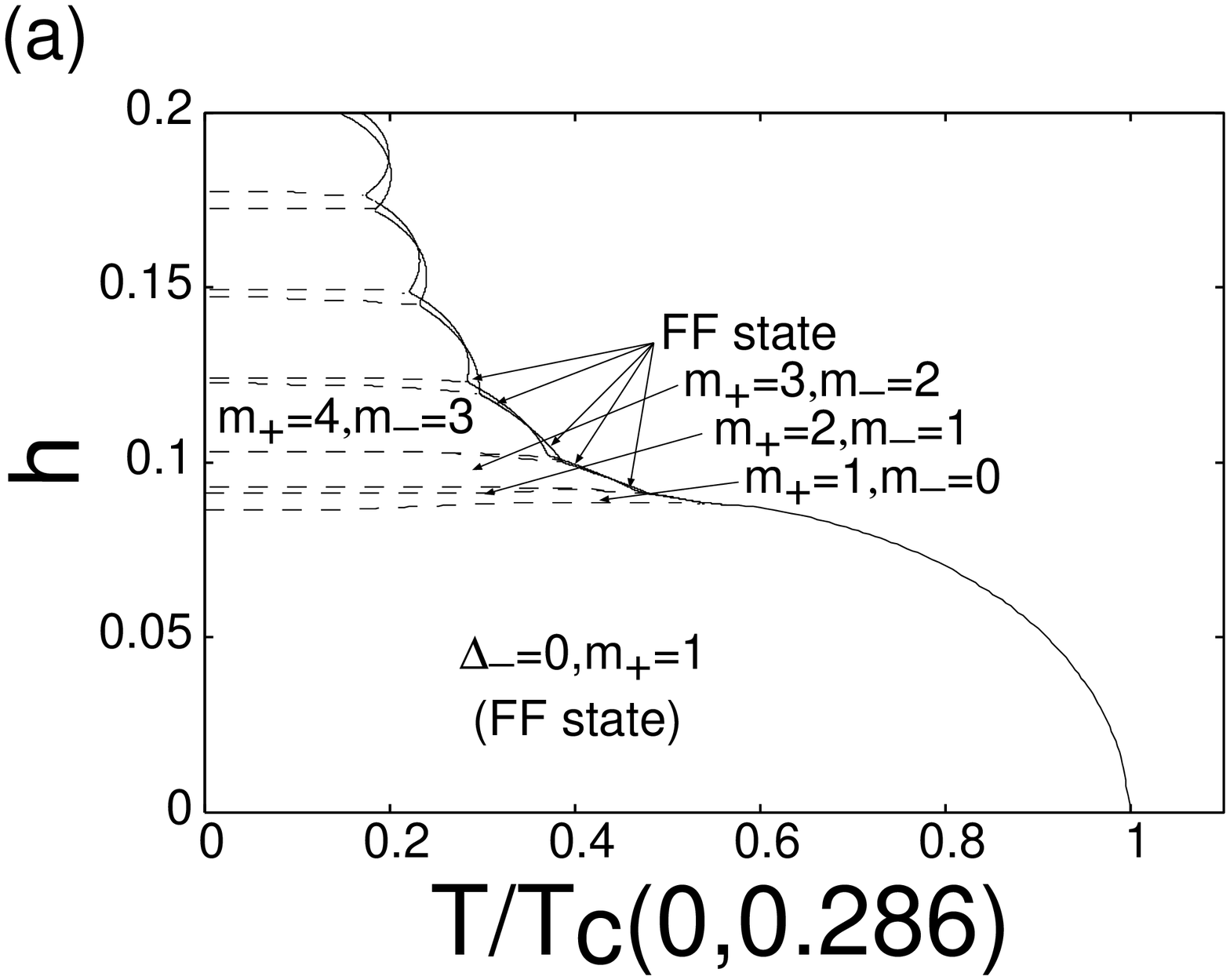}
    \end{minipage}
    \begin{minipage}{0.5\hsize}
      \includegraphics[width=70mm]{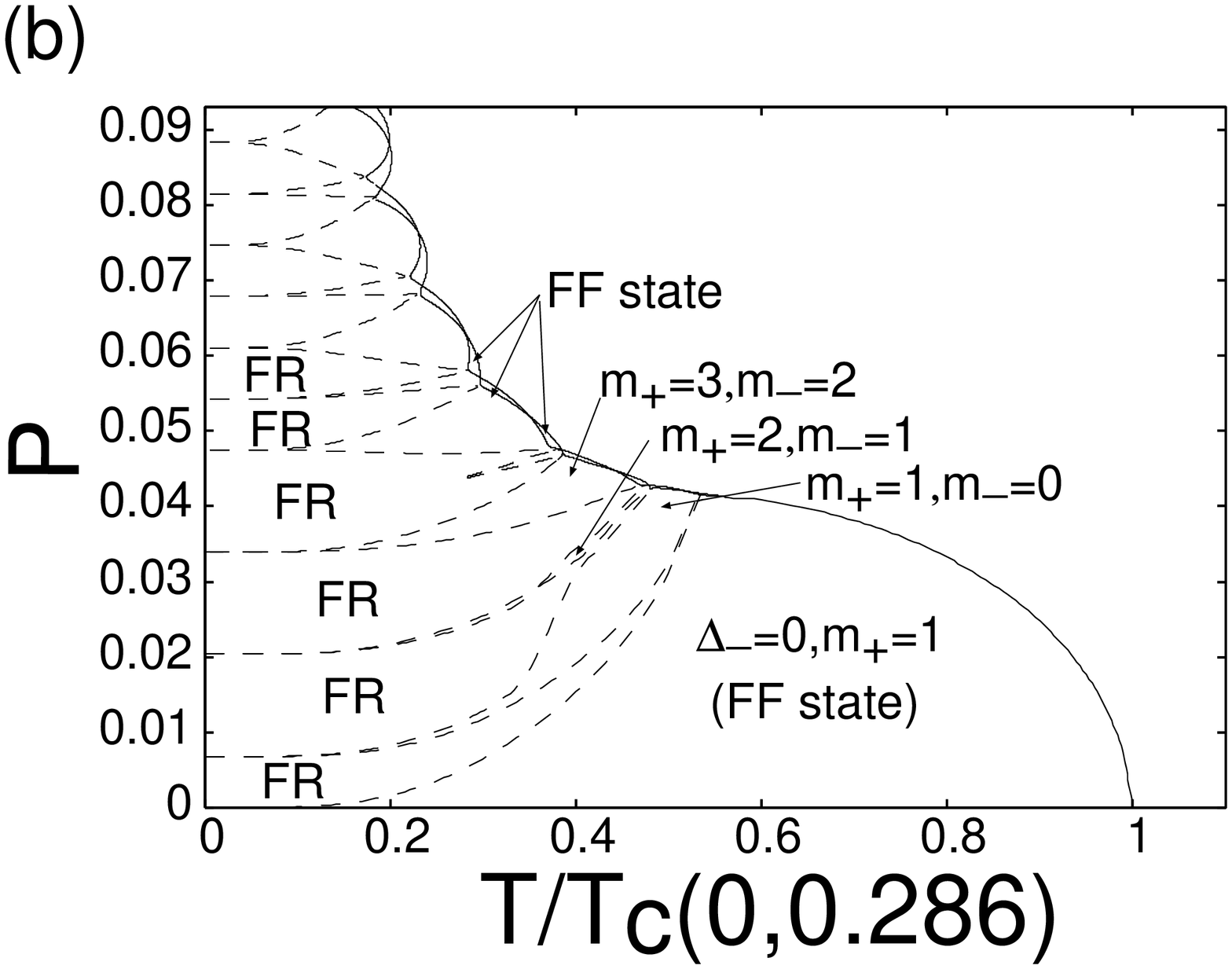}
    \end{minipage}
  \end{tabular}
  \begin{tabular}{cc}
    \begin{minipage}{0.5\hsize}
      \includegraphics[width=70mm]{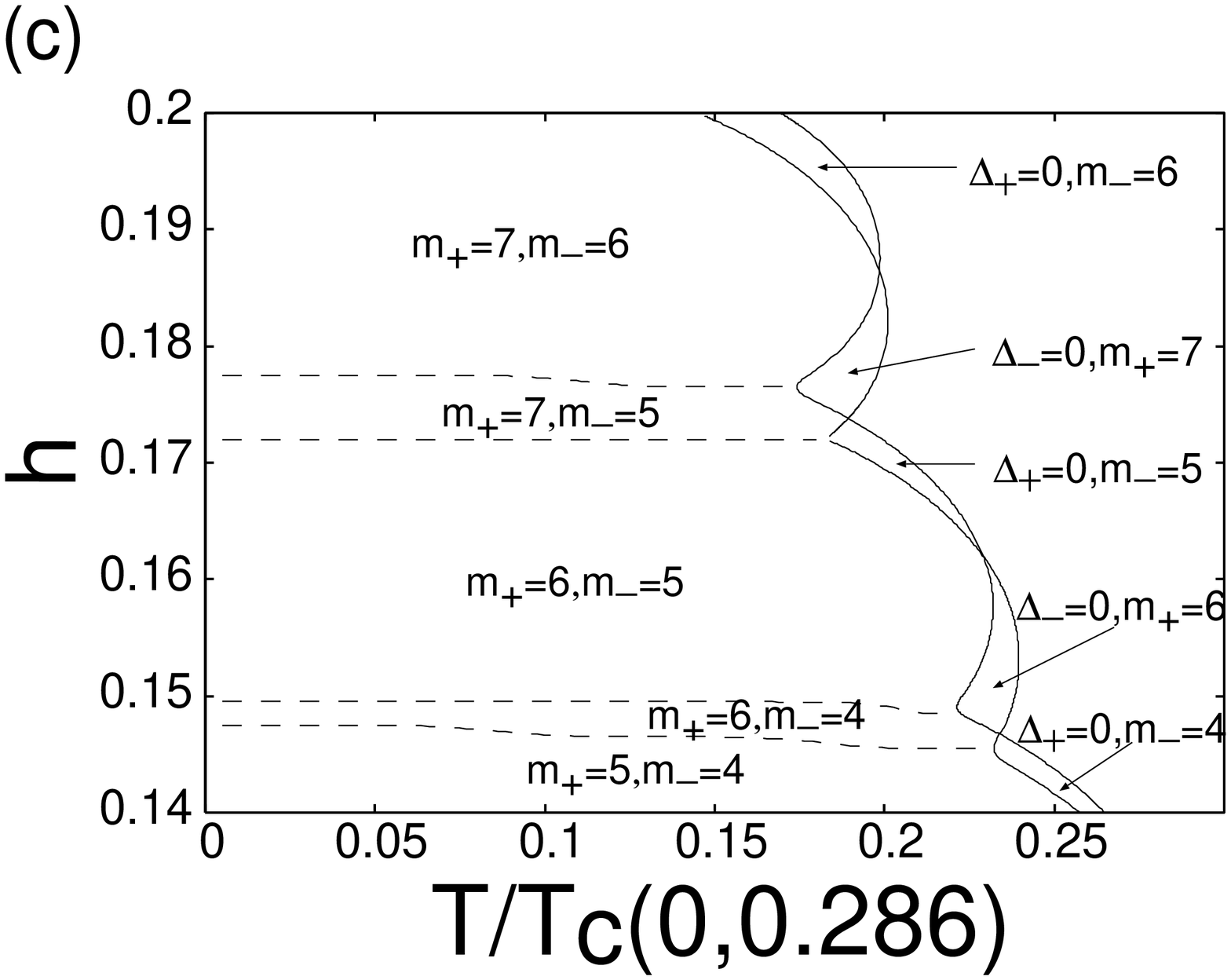}
    \end{minipage}
    \begin{minipage}{0.5\hsize}
      \includegraphics[width=70mm]{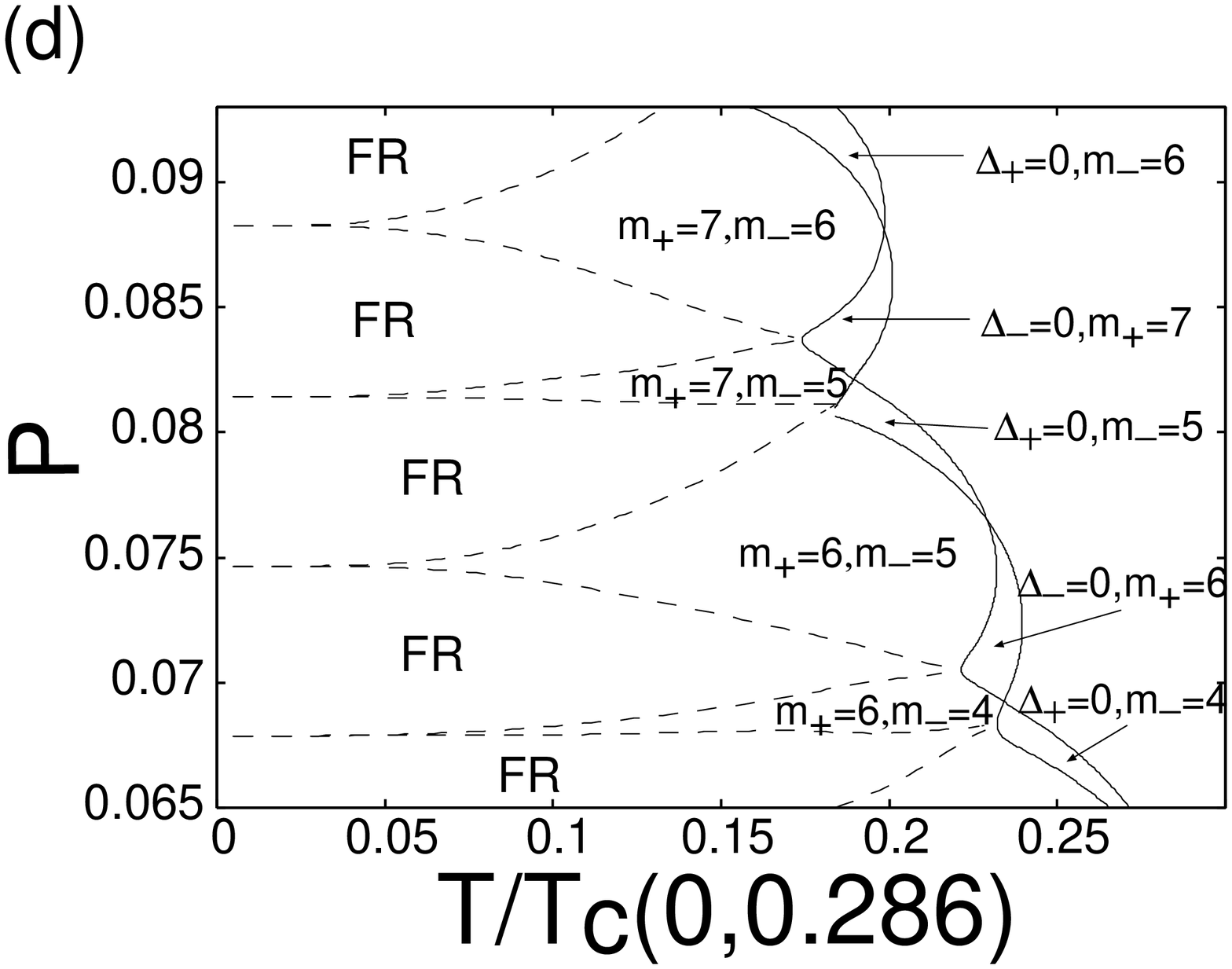}
    \end{minipage}
  \end{tabular}
  \caption{(a) $T-h$ and (b) $T-P$ phase diagram for $\Phi/\Phi_0=0.286$. 
    Scaled-up figures of (a) and (b) are shown in (c) and (d), respectively.
    \label{fig:fig9}}
\end{figure*}

Finally, we discuss the phase diagram of the rotating Fermi superfluid 
gases in the $T-h$ and $T-P$ plane. 
We fix the rotation $\Phi/\Phi_0=0.286$ in 
Fig.~\ref{fig:fig9}. 
For this parameter the BCS state is changed to the vortex state with 
vorticity $m=1$. 
More important finding is that the half quantum vortex state 
[for example, $(m_+,m_-)=(1,0)$] is 
induced in a certain range of the population imbalance. 
Therefore, the flux of mass iteratively changes as $1 \rightarrow \frac{1}{2} 
\rightarrow  1 \rightarrow  \frac{1}{2}$ with increasing the population 
imbalance. 
Another point is that the FF state is stabilized near the critical 
temperature, as studied in section \ref{sec:sec3a}.  
In order to illuminate these features, we show the phase diagram scaled-up
in the large population imbalance (magnetic field) region. 
We see that the population imbalance changes the vorticity on a 
sequence of phase transition lines. 
These features are understood on the basis of the mesoscopic effect 
discussed in section \ref{sec:sec3c}. 

For a rotation smaller than $\Phi/\Phi_0=0.286$,   
where the vortex is not nucleated in the balanced gas, 
a half quantum vortex can be induced by the population imbalance
(compare Fig.~\ref{fig:fig2}(a) with Fig.~\ref{fig:fig2}(b)). 
Then, the gas begins to rotate by increasing the population imbalance, 
if the angular momentum is not conserved for some reasons. 
This intriguing phenomenon would be a signature of the FFLO superfluid gas. 

\section{Summary and discussion}

We have studied effects of rotation on the FFLO superfluid state 
of imbalanced Fermi gases loaded on a quasi-one-dimensional 
toroidal trap. 
We found several novel phase transitions induced by the rotation 
and population imbalance. 
First, the FF state is stabilized in the rotating gases 
near the critical temperature, although this state is unstable 
against the LO state in the gases at rest. 
This means that the ``giant vortex'' appears in the imbalanced gases 
near $T_{\rm c}$ because the FF state on the ring is regarded as 
a vortex state. 
We found that the vorticity shows a nonmonotonic and intriguing dependence 
on the population imbalance. 
Second, the FF state changes to the intermediate state through the 
second order phase transition at $T=T_{\rm c2} < T_{\rm c}$. 
As decreasing the temperature below $T_{\rm c2}$, the crossover occurs 
from the intermediate state to the LO state. 
We point out that the FF, LO, and intermediate phases are 
distinguished by the flux of mass and local polarization. 
Third, we showed a sequence of quantum phase transitions with increasing
the rotation. The FFLO state with an integer vortex changes to the 
half quantum vortex state. 
Then, the flux of mass is half quantized (Fig.~\ref{fig:fig3}). 
The half quantum vortex state arises from the multi-component 
order parameters of the FFLO state, and therefore the half quantized 
flux would be a clear evidence for the FFLO superfluid state. 
The spatial profile of the order parameter in the half quantum vortex 
state is described by the product  
$\Delta_{\rm FF} \times \Delta_{\rm LO}$ where $\Delta_{\rm FF}$ and 
$\Delta_{\rm LO}$ are the order parameters of FF and LO states, 
respectively. 

These results are obtained on the basis of the BdG equation. 
Since the BdG equation is a mean field approximation 
it breaks down in the purely one-dimensional system. 
It is well known that the long-range order of 
superfluidity/superconductivity is fragile against 
the fluctuation in one dimension.  
We adopt the one-dimensional model for simplicity, however 
our interests are focused on the quasi-one-dimensional 
superfluid which is realized in the toroidal 
trap~\cite{PhysRevB.80.220510,ryu:260401,Sherlock}. 
It has been shown that a weak three-dimensional coupling gives rise to 
the long range order~\cite{parish:250403,zhao:063605,PhysRevA.84.033609}. 
The BdG equation is qualitatively appropriate for a weak 
attractive interaction in this case. 
The correlation effect beyond the mean field approximation 
plays important roles for a strong attractive interaction.  
Therefore, our study basically focuses on the weak coupling BCS region. 
However, it has been shown that the FFLO superfluid state 
is robust against the strong attractive interaction 
in the unitary regime~\cite{bulgac:215301,PhysRevB.80.220510}. 
This indicates that the novel FFLO phases studied in this paper 
are robust in the BCS-BEC crossover region. 

We note that a fingerprint of the FFLO phases appears 
in the purely one-dimensional system~\cite{Liao}, 
although the long range order is suppressed there. 
For instance, the free energy shows a characteristic 
periodicity with respect to the phase $\Phi$ as shown in 
Ref.~\cite{PhysRevLett.70.978}. 
We expect that the free energy of the imbalanced gases shows 
an oscillation with respect to the phase $\Phi$, and the period 
is a half of the BCS state.
This oscillation is a fingerprint of the half quantum vortex state 
in the FFLO state. 
It will be interesting to see this periodicity with use of the 
accurate numerical and/or analytical methods for one-dimensional models.  

Three comments are given on the related theoretical works. 
First, the same model as in Eq.~(\ref{eq:eq1}) was solved in the recent 
paper~\cite{PhysRevB.81.014518} on the basis of the BdG equations. 
However, they missed most of the novel FFLO phases induced by the 
rotation, except for the vortex state in the small imbalance region. 
We showed $T-h$ and $T-P$ phase diagrams in which the FF state with 
giant vortex, half quantum vortex state, and intermediate state are 
stabilized. 
These phases were not shown in Ref.~\cite{PhysRevB.81.014518}. 

Second, a superfluid state similar to the A-FFLO state was investigated 
in the optical lattice by Chen {\it et al.}~\cite{PhysRevB.79.054512}.
The response to the rotation is quite different from our case, 
because the rotation symmetry 
is broken by the optical lattice potential. 
For example, the FF state and the intermediate state are suppressed 
because the degeneracy of the FF states is lifted. 
In other words, the FFLO state in the optical lattice does not have 
an internal degree of freedom in contrast to the continuum gases. 
It is interesting to study the effect of broken rotation symmetry 
by tuning the aspect ratio~\cite{Sherlock}.
Then, other FFLO states may be stabilized by the rotation. 

Third, the rotating FFLO phases studied in this paper are regarded as 
an analog of the superconducting states in the non-centrosymmetric 
systems~\cite{Agterberg_helical}. 
The degeneracy of the FF phases is lifted in the 
latter case by the anti-symmetric spin-orbit coupling arising from 
the broken inversion symmetry. 
The helical superconducting state is an analog of the FF state 
in our case, while the stripe phase corresponds to the intermediate 
state. 
However, these non-centrosymmetric superconducting phases 
cannot be realized because the orbital de-pairing effect arising 
from the magnetic field affects the superconducting state. 
In contrast, cold Fermi gases are ideal system for the study of 
FFLO phases thanks to their high controllability. 

A ring trap has been realized for Bose gases~\cite{ryu:260401,Sherlock}, 
and produces many intriguing topics. 
We expect that the realization of Fermi superfluid on the ring trap will 
open a new research field of cold atom gases, as the development of 
sophisticated trap potentials triggered novel ideas. 
This paper shows that the rotating FFLO superfluid phases in the 
imbalanced Fermi gases will be one of the goals. 
We hope that active investigations will elucidate such novel phases. 

\begin{acknowledgments}
The authors are grateful to G. Marmorini, S. Tsuchiya, and Y. Ono 
for fruitful discussions. 
T. Y. thanks  Y. Yamakawa for variable comment on the numerical calculations. 
This work was supported by a Grant-in-Aid for Scientific Research 
on Innovative Areas ``Heavy Electrons'' (No. 21102506) from MEXT, Japan. 
It was also supported by a Grant-in-Aid for 
Young Scientists (B) (No. 20740187) from JSPS. 
Numerical computation in this work was partly carried out 
at the Yukawa Institute Computer Facility. 
\end{acknowledgments}

\appendix
\section{Derivation of the Hamiltonian}
In Appendix A, we derive the lattice Hamiltonian 
in Eq.~(\ref{eq:eq1}) from the continuum model. 
The Hamiltonian of continuum one-dimensional gases is described  
in the rotating coordinate as 
\begin{eqnarray}
H&=&\sum_\sigma\int dx\hat{\psi}_\sigma^\dagger(x)\biggl(-\frac{1}{2M}\frac{\partial^2}{\partial x^2}
-\Omega \hat{L}_z\biggr)\hat{\psi}_\sigma(x) \nonumber \\
&&-|U|\int dx\hat{\psi}_\uparrow^\dagger(x)\hat{\psi}_\downarrow^\dagger(x)
\hat{\psi}_\downarrow(x)\hat{\psi}_\uparrow(x), \nonumber \\
&=&H_0-|U|\int dx\hat{\psi}_\uparrow^\dagger(x)\hat{\psi}_\downarrow^\dagger(x)
\hat{\psi}_\downarrow(x)\hat{\psi}_\uparrow(x),
\end{eqnarray}
where $\hat{\psi}_\sigma^\dagger(x)$ and $\hat{\psi}_\sigma(x)$ are the fermionic creation and annihilation operators, respectively, 
and $H_0$ is the single-particle Hamiltonian.

The coordinate along the ring $x$ is given in term of 
the angle $\theta$ and the radius of ring $R$. 
Then, the {\it z}-component of the angular momentum $\hat{L}_{\rm z}$ 
is described as 
$\hat{L}_z=-i\frac{\partial}{\partial\theta}=-iR\frac{\partial}{\partial x}$.

With use of the Fourier transformation, 
the single-particle Hamiltonian $H_0$ is obtained as  
\begin{eqnarray}
H_0=\sum_{k,\sigma}\biggl(\frac{k^2}{2M}-R\Omega k\biggr)\hat{c}_{k\sigma}^\dagger \hat{c}_{k\sigma}.
\end{eqnarray}
Thus, the single particle energy is obtained as 
$\varepsilon(k) = \frac{k^2}{2M}-R\Omega k $.  
It is understood that the Fermi momentum is shifted 
to the positive direction of $k$ for $\Omega > 0$.  
On the other hand, the single particle energy in the lattice Hamiltonian 
(Eq.~(\ref{eq:eq1})) is obtained as 
\begin{eqnarray}
\varepsilon(k)=-2t\cos(k-\Phi). 
\end{eqnarray}
This is approximated in the continuum limit $k, \Phi \ll 1$ as   
\begin{eqnarray}
\varepsilon(k)\simeq 
t(\Phi^2-2)-2t\Phi k+tk^2.
\end{eqnarray}
Hence, we obtain the parameters of the lattice Hamiltonian 
Eq.~(\ref{eq:eq1}) by which the continuum gases are 
approximately modeled, 
\begin{eqnarray}
t\leftrightarrow \frac{1}{2M},\  2t\Phi\leftrightarrow R\Omega.
\end{eqnarray}

\section{Periodicity with respect to $\Phi$} 

With use of the gauge transformation, 
\begin{eqnarray}
\hat{c}_{j\sigma}\rightarrow {\rm e}^{-i\Phi j}\hat{c}_{j\sigma},
\end{eqnarray}
the hopping term is transformed as 
\begin{eqnarray}
-t\hat{c}^\dagger_{j+1\sigma}\hat{c}_{j\sigma}\rightarrow -t{\rm e}^{i\Phi}\hat{c}^\dagger_{j+1\sigma}\hat{c}_{j\sigma}.
\end{eqnarray}
Since we adopt the periodic boundary condition, 
the phase $\Phi$ should be 
\begin{eqnarray}
\Phi_n=\frac{2\pi}{N_{\rm L}}n,
\end{eqnarray}
where $n$ is an integer. 
This gauge transformation changes the Hamiltonian for the gas at rest 
to that for the rotating gas. 
This means that the physical quantities are periodic in $\Phi$ 
with the period $2\pi/N_{\rm L} = \Phi_0$. 
Hence we show the results for $0 \leq \Phi/\Phi_0 < 1$ 
in Figs.~\ref{fig:fig2} and \ref{fig:fig7}. 

\bibliography{reference}

\begin{thebibliography}{55}%
\makeatletter
\providecommand \@ifxundefined [1]{%
 \@ifx{#1\undefined}
}%
\providecommand \@ifnum [1]{%
 \ifnum #1\expandafter \@firstoftwo
 \else \expandafter \@secondoftwo
 \fi
}%
\providecommand \@ifx [1]{%
 \ifx #1\expandafter \@firstoftwo
 \else \expandafter \@secondoftwo
 \fi
}%
\providecommand \natexlab [1]{#1}%
\providecommand \enquote  [1]{``#1''}%
\providecommand \bibnamefont  [1]{#1}%
\providecommand \bibfnamefont [1]{#1}%
\providecommand \citenamefont [1]{#1}%
\providecommand \href@noop [0]{\@secondoftwo}%
\providecommand \href [0]{\begingroup \@sanitize@url \@href}%
\providecommand \@href[1]{\@@startlink{#1}\@@href}%
\providecommand \@@href[1]{\endgroup#1\@@endlink}%
\providecommand \@sanitize@url [0]{\catcode `\\12\catcode `\$12\catcode
  `\&12\catcode `\#12\catcode `\^12\catcode `\_12\catcode `\%12\relax}%
\providecommand \@@startlink[1]{}%
\providecommand \@@endlink[0]{}%
\providecommand \url  [0]{\begingroup\@sanitize@url \@url }%
\providecommand \@url [1]{\endgroup\@href {#1}{\urlprefix }}%
\providecommand \urlprefix  [0]{URL }%
\providecommand \Eprint [0]{\href }%
\providecommand \doibase [0]{http://dx.doi.org/}%
\providecommand \selectlanguage [0]{\@gobble}%
\providecommand \bibinfo  [0]{\@secondoftwo}%
\providecommand \bibfield  [0]{\@secondoftwo}%
\providecommand \translation [1]{[#1]}%
\providecommand \BibitemOpen [0]{}%
\providecommand \bibitemStop [0]{}%
\providecommand \bibitemNoStop [0]{.\EOS\space}%
\providecommand \EOS [0]{\spacefactor3000\relax}%
\providecommand \BibitemShut  [1]{\csname bibitem#1\endcsname}%
\let\auto@bib@innerbib\@empty
\bibitem [{\citenamefont {Partridge}\ \emph {et~al.}(2006)\citenamefont
  {Partridge}, \citenamefont {Li}, \citenamefont {Kamar}, \citenamefont
  {Liao},\ and\ \citenamefont {Hulet}}]{Partridge27012006}%
  \BibitemOpen
  \bibfield  {author} {\bibinfo {author} {\bibfnamefont {G.~B.}\ \bibnamefont
  {Partridge}}, \bibinfo {author} {\bibfnamefont {W.}~\bibnamefont {Li}},
  \bibinfo {author} {\bibfnamefont {R.~I.}\ \bibnamefont {Kamar}}, \bibinfo
  {author} {\bibfnamefont {Y.-a.}\ \bibnamefont {Liao}}, \ and\ \bibinfo
  {author} {\bibfnamefont {R.~G.}\ \bibnamefont {Hulet}},\ }\href {\doibase
  10.1126/science.1122876} {\bibfield  {journal} {\bibinfo  {journal}
  {Science}\ }\textbf {\bibinfo {volume} {311}},\ \bibinfo {pages} {503}
  (\bibinfo {year} {2006})}\BibitemShut {NoStop}%
\bibitem [{\citenamefont {Zwierlein}\ \emph {et~al.}(2006)\citenamefont
  {Zwierlein}, \citenamefont {Schirotzek}, \citenamefont {Schunck},\ and\
  \citenamefont {Ketterle}}]{Zwierlein27012006}%
  \BibitemOpen
  \bibfield  {author} {\bibinfo {author} {\bibfnamefont {M.~W.}\ \bibnamefont
  {Zwierlein}}, \bibinfo {author} {\bibfnamefont {A.}~\bibnamefont
  {Schirotzek}}, \bibinfo {author} {\bibfnamefont {C.~H.}\ \bibnamefont
  {Schunck}}, \ and\ \bibinfo {author} {\bibfnamefont {W.}~\bibnamefont
  {Ketterle}},\ }\href {\doibase 10.1126/science.1122318} {\bibfield  {journal}
  {\bibinfo  {journal} {Science}\ }\textbf {\bibinfo {volume} {311}},\ \bibinfo
  {pages} {492} (\bibinfo {year} {2006})}\BibitemShut {NoStop}%
\bibitem [{\citenamefont {Giorgini}\ \emph {et~al.}(2008)\citenamefont
  {Giorgini}, \citenamefont {Pitaevskii},\ and\ \citenamefont
  {Stringari}}]{RevModPhys.80.1215}%
  \BibitemOpen
  \bibfield  {author} {\bibinfo {author} {\bibfnamefont {S.}~\bibnamefont
  {Giorgini}}, \bibinfo {author} {\bibfnamefont {L.~P.}\ \bibnamefont
  {Pitaevskii}}, \ and\ \bibinfo {author} {\bibfnamefont {S.}~\bibnamefont
  {Stringari}},\ }\href {\doibase 10.1103/RevModPhys.80.1215} {\bibfield
  {journal} {\bibinfo  {journal} {Rev. Mod. Phys.}\ }\textbf {\bibinfo {volume}
  {80}},\ \bibinfo {pages} {1215} (\bibinfo {year} {2008})}\BibitemShut
  {NoStop}%
\bibitem [{\citenamefont {Fulde}\ and\ \citenamefont
  {Ferrell}(1964)}]{PhysRev.135.A550}%
  \BibitemOpen
  \bibfield  {author} {\bibinfo {author} {\bibfnamefont {P.}~\bibnamefont
  {Fulde}}\ and\ \bibinfo {author} {\bibfnamefont {R.~A.}\ \bibnamefont
  {Ferrell}},\ }\href {\doibase 10.1103/PhysRev.135.A550} {\bibfield  {journal}
  {\bibinfo  {journal} {Phys. Rev.}\ }\textbf {\bibinfo {volume} {135}},\
  \bibinfo {pages} {A550} (\bibinfo {year} {1964})}\BibitemShut {NoStop}%
\bibitem [{\citenamefont {Larkin}\ and\ \citenamefont
  {Ovchinnikov}(1965)}]{SovPhysJETP.20.762}%
  \BibitemOpen
  \bibfield  {author} {\bibinfo {author} {\bibfnamefont {A.~I.}\ \bibnamefont
  {Larkin}}\ and\ \bibinfo {author} {\bibfnamefont {Y.~N.}\ \bibnamefont
  {Ovchinnikov}},\ }\href@noop {} {\bibfield  {journal} {\bibinfo  {journal}
  {Sov. Phys. JETP}\ }\textbf {\bibinfo {volume} {20}},\ \bibinfo {pages} {762}
  (\bibinfo {year} {1965})}\BibitemShut {NoStop}%
\bibitem [{\citenamefont {Liao}\ \emph {et~al.}(2010)\citenamefont {Liao},
  \citenamefont {Rittner}, \citenamefont {Paprotta}, \citenamefont {Li},
  \citenamefont {Partridge}, \citenamefont {Hulet}, \citenamefont {Baur},\ and\
  \citenamefont {Mueller}}]{Liao}%
  \BibitemOpen
  \bibfield  {author} {\bibinfo {author} {\bibfnamefont {Y.-a.}\ \bibnamefont
  {Liao}}, \bibinfo {author} {\bibfnamefont {A.~S.~C.}\ \bibnamefont
  {Rittner}}, \bibinfo {author} {\bibfnamefont {T.}~\bibnamefont {Paprotta}},
  \bibinfo {author} {\bibfnamefont {W.}~\bibnamefont {Li}}, \bibinfo {author}
  {\bibfnamefont {G.~B.}\ \bibnamefont {Partridge}}, \bibinfo {author}
  {\bibfnamefont {R.~G.}\ \bibnamefont {Hulet}}, \bibinfo {author}
  {\bibfnamefont {S.~K.}\ \bibnamefont {Baur}}, \ and\ \bibinfo {author}
  {\bibfnamefont {E.~J.}\ \bibnamefont {Mueller}},\ }\href@noop {} {\bibfield
  {journal} {\bibinfo  {journal} {Nature (London)}\ }\textbf {\bibinfo {volume}
  {467}},\ \bibinfo {pages} {567} (\bibinfo {year} {2010})}\BibitemShut
  {NoStop}%
\bibitem [{\citenamefont {Radovan}\ \emph {et~al.}(2003)\citenamefont
  {Radovan}, \citenamefont {Fortune}, \citenamefont {Murphy}, \citenamefont
  {Hannahs}, \citenamefont {Palm}, \citenamefont {Tozer},\ and\ \citenamefont
  {Hall}}]{Nature.425.51}%
  \BibitemOpen
  \bibfield  {author} {\bibinfo {author} {\bibfnamefont {H.~A.}\ \bibnamefont
  {Radovan}}, \bibinfo {author} {\bibfnamefont {N.~A.}\ \bibnamefont
  {Fortune}}, \bibinfo {author} {\bibfnamefont {T.~P.}\ \bibnamefont {Murphy}},
  \bibinfo {author} {\bibfnamefont {S.~T.}\ \bibnamefont {Hannahs}}, \bibinfo
  {author} {\bibfnamefont {E.~C.}\ \bibnamefont {Palm}}, \bibinfo {author}
  {\bibfnamefont {S.~W.}\ \bibnamefont {Tozer}}, \ and\ \bibinfo {author}
  {\bibfnamefont {D.}~\bibnamefont {Hall}},\ }\href@noop {} {\bibfield
  {journal} {\bibinfo  {journal} {Nature (London)}\ }\textbf {\bibinfo {volume}
  {425}},\ \bibinfo {pages} {51} (\bibinfo {year} {2003})}\BibitemShut
  {NoStop}%
\bibitem [{\citenamefont {Bianchi}\ \emph {et~al.}(2003)\citenamefont
  {Bianchi}, \citenamefont {Movshovich}, \citenamefont {Capan}, \citenamefont
  {Pagliuso},\ and\ \citenamefont {Sarrao}}]{Bianchi_FFLO}%
  \BibitemOpen
  \bibfield  {author} {\bibinfo {author} {\bibfnamefont {A.}~\bibnamefont
  {Bianchi}}, \bibinfo {author} {\bibfnamefont {R.}~\bibnamefont {Movshovich}},
  \bibinfo {author} {\bibfnamefont {C.}~\bibnamefont {Capan}}, \bibinfo
  {author} {\bibfnamefont {P.~G.}\ \bibnamefont {Pagliuso}}, \ and\ \bibinfo
  {author} {\bibfnamefont {J.~L.}\ \bibnamefont {Sarrao}},\ }\href {\doibase
  10.1103/PhysRevLett.91.187004} {\bibfield  {journal} {\bibinfo  {journal}
  {Phys. Rev. Lett.}\ }\textbf {\bibinfo {volume} {91}},\ \bibinfo {pages}
  {187004} (\bibinfo {year} {2003})}\BibitemShut {NoStop}%
\bibitem [{\citenamefont {Matsuda}\ and\ \citenamefont
  {Shimahara}(2007)}]{JPSJ.76.051005}%
  \BibitemOpen
  \bibfield  {author} {\bibinfo {author} {\bibfnamefont {Y.}~\bibnamefont
  {Matsuda}}\ and\ \bibinfo {author} {\bibfnamefont {H.}~\bibnamefont
  {Shimahara}},\ }\href {\doibase 10.1143/JPSJ.76.051005} {\bibfield  {journal}
  {\bibinfo  {journal} {J. Phys. Soc. Jpn.}\ }\textbf {\bibinfo {volume}
  {76}},\ \bibinfo {pages} {051005} (\bibinfo {year} {2007})}\BibitemShut
  {NoStop}%
\bibitem [{\citenamefont {Kenzelmann}\ \emph {et~al.}(2008)\citenamefont
  {Kenzelmann}, \citenamefont {Str\"{a}ssle}, \citenamefont {Niedermayer},
  \citenamefont {Sigrist}, \citenamefont {Padmanabhan}, \citenamefont
  {Zolliker}, \citenamefont {Bianchi}, \citenamefont {Movshovich},
  \citenamefont {Bauer}, \citenamefont {Sarrao},\ and\ \citenamefont
  {Thompson}}]{Kenzelmann19092008}%
  \BibitemOpen
  \bibfield  {author} {\bibinfo {author} {\bibfnamefont {M.}~\bibnamefont
  {Kenzelmann}}, \bibinfo {author} {\bibfnamefont {T.}~\bibnamefont
  {Str\"{a}ssle}}, \bibinfo {author} {\bibfnamefont {C.}~\bibnamefont
  {Niedermayer}}, \bibinfo {author} {\bibfnamefont {M.}~\bibnamefont
  {Sigrist}}, \bibinfo {author} {\bibfnamefont {B.}~\bibnamefont
  {Padmanabhan}}, \bibinfo {author} {\bibfnamefont {M.}~\bibnamefont
  {Zolliker}}, \bibinfo {author} {\bibfnamefont {A.~D.}\ \bibnamefont
  {Bianchi}}, \bibinfo {author} {\bibfnamefont {R.}~\bibnamefont {Movshovich}},
  \bibinfo {author} {\bibfnamefont {E.~D.}\ \bibnamefont {Bauer}}, \bibinfo
  {author} {\bibfnamefont {J.~L.}\ \bibnamefont {Sarrao}}, \ and\ \bibinfo
  {author} {\bibfnamefont {J.~D.}\ \bibnamefont {Thompson}},\ }\href {\doibase
  10.1126/science.1161818} {\bibfield  {journal} {\bibinfo  {journal}
  {Science}\ }\textbf {\bibinfo {volume} {321}},\ \bibinfo {pages} {1652}
  (\bibinfo {year} {2008})}\BibitemShut {NoStop}%
\bibitem [{\citenamefont {Ikeda}(2007{\natexlab{a}})}]{PhysRevB.76.134504}%
  \BibitemOpen
  \bibfield  {author} {\bibinfo {author} {\bibfnamefont {R.}~\bibnamefont
  {Ikeda}},\ }\href {\doibase 10.1103/PhysRevB.76.134504} {\bibfield  {journal}
  {\bibinfo  {journal} {Phys. Rev. B}\ }\textbf {\bibinfo {volume} {76}},\
  \bibinfo {pages} {134504} (\bibinfo {year} {2007}{\natexlab{a}})}\BibitemShut
  {NoStop}%
\bibitem [{\citenamefont {Ikeda}(2007{\natexlab{b}})}]{PhysRevB.76.054517}%
  \BibitemOpen
  \bibfield  {author} {\bibinfo {author} {\bibfnamefont {R.}~\bibnamefont
  {Ikeda}},\ }\href {\doibase 10.1103/PhysRevB.76.054517} {\bibfield  {journal}
  {\bibinfo  {journal} {Phys. Rev. B}\ }\textbf {\bibinfo {volume} {76}},\
  \bibinfo {pages} {054517} (\bibinfo {year} {2007}{\natexlab{b}})}\BibitemShut
  {NoStop}%
\bibitem [{\citenamefont {Agterberg}\ \emph {et~al.}(2009)\citenamefont
  {Agterberg}, \citenamefont {Sigrist},\ and\ \citenamefont
  {Tsunetsugu}}]{PhysRevLett.102.207004}%
  \BibitemOpen
  \bibfield  {author} {\bibinfo {author} {\bibfnamefont {D.~F.}\ \bibnamefont
  {Agterberg}}, \bibinfo {author} {\bibfnamefont {M.}~\bibnamefont {Sigrist}},
  \ and\ \bibinfo {author} {\bibfnamefont {H.}~\bibnamefont {Tsunetsugu}},\
  }\href {\doibase 10.1103/PhysRevLett.102.207004} {\bibfield  {journal}
  {\bibinfo  {journal} {Phys. Rev. Lett.}\ }\textbf {\bibinfo {volume} {102}},\
  \bibinfo {pages} {207004} (\bibinfo {year} {2009})}\BibitemShut {NoStop}%
\bibitem [{\citenamefont {Aperis}\ \emph {et~al.}(2010)\citenamefont {Aperis},
  \citenamefont {Varelogiannis},\ and\ \citenamefont
  {Littlewood}}]{aperis2010}%
  \BibitemOpen
  \bibfield  {author} {\bibinfo {author} {\bibfnamefont {A.}~\bibnamefont
  {Aperis}}, \bibinfo {author} {\bibfnamefont {G.}~\bibnamefont
  {Varelogiannis}}, \ and\ \bibinfo {author} {\bibfnamefont {P.~B.}\
  \bibnamefont {Littlewood}},\ }\href {\doibase 10.1103/PhysRevLett.104.216403}
  {\bibfield  {journal} {\bibinfo  {journal} {Phys. Rev. Lett.}\ }\textbf
  {\bibinfo {volume} {104}},\ \bibinfo {pages} {216403} (\bibinfo {year}
  {2010})}\BibitemShut {NoStop}%
\bibitem [{\citenamefont {Yanase}\ and\ \citenamefont
  {Sigrist}(2009)}]{JPSJ.78.114715}%
  \BibitemOpen
  \bibfield  {author} {\bibinfo {author} {\bibfnamefont {Y.}~\bibnamefont
  {Yanase}}\ and\ \bibinfo {author} {\bibfnamefont {M.}~\bibnamefont
  {Sigrist}},\ }\href {\doibase 10.1143/JPSJ.78.114715} {\bibfield  {journal}
  {\bibinfo  {journal} {J. Phys. Soc. Jpn.}\ }\textbf {\bibinfo {volume}
  {78}},\ \bibinfo {pages} {114715} (\bibinfo {year} {2009})}\BibitemShut
  {NoStop}%
\bibitem [{\citenamefont {Casalbuoni}\ and\ \citenamefont
  {Nardulli}(2004)}]{RevModPhys.76.263}%
  \BibitemOpen
  \bibfield  {author} {\bibinfo {author} {\bibfnamefont {R.}~\bibnamefont
  {Casalbuoni}}\ and\ \bibinfo {author} {\bibfnamefont {G.}~\bibnamefont
  {Nardulli}},\ }\href {\doibase 10.1103/RevModPhys.76.263} {\bibfield
  {journal} {\bibinfo  {journal} {Rev. Mod. Phys.}\ }\textbf {\bibinfo {volume}
  {76}},\ \bibinfo {pages} {263} (\bibinfo {year} {2004})}\BibitemShut
  {NoStop}%
\bibitem [{\citenamefont {Sheehy}\ and\ \citenamefont
  {Radzihovsky}(2006)}]{sheehy:060401}%
  \BibitemOpen
  \bibfield  {author} {\bibinfo {author} {\bibfnamefont {D.~E.}\ \bibnamefont
  {Sheehy}}\ and\ \bibinfo {author} {\bibfnamefont {L.}~\bibnamefont
  {Radzihovsky}},\ }\href {\doibase 10.1103/PhysRevLett.96.060401} {\bibfield
  {journal} {\bibinfo  {journal} {Phys. Rev. Lett.}\ }\textbf {\bibinfo
  {volume} {96}},\ \bibinfo {eid} {060401} (\bibinfo {year}
  {2006})}\BibitemShut {NoStop}%
\bibitem [{\citenamefont {Yoshida}\ and\ \citenamefont
  {Yip}(2007)}]{yoshida:063601}%
  \BibitemOpen
  \bibfield  {author} {\bibinfo {author} {\bibfnamefont {N.}~\bibnamefont
  {Yoshida}}\ and\ \bibinfo {author} {\bibfnamefont {S.-K.}\ \bibnamefont
  {Yip}},\ }\href {\doibase 10.1103/PhysRevA.75.063601} {\bibfield  {journal}
  {\bibinfo  {journal} {Phys. Rev. A}\ }\textbf {\bibinfo {volume} {75}},\
  \bibinfo {eid} {063601} (\bibinfo {year} {2007})}\BibitemShut {NoStop}%
\bibitem [{\citenamefont {Castorina}\ \emph {et~al.}(2005)\citenamefont
  {Castorina}, \citenamefont {Grasso}, \citenamefont {Oertel}, \citenamefont
  {Urban},\ and\ \citenamefont {Zappal\`a}}]{PhysRevA.72.025601}%
  \BibitemOpen
  \bibfield  {author} {\bibinfo {author} {\bibfnamefont {P.}~\bibnamefont
  {Castorina}}, \bibinfo {author} {\bibfnamefont {M.}~\bibnamefont {Grasso}},
  \bibinfo {author} {\bibfnamefont {M.}~\bibnamefont {Oertel}}, \bibinfo
  {author} {\bibfnamefont {M.}~\bibnamefont {Urban}}, \ and\ \bibinfo {author}
  {\bibfnamefont {D.}~\bibnamefont {Zappal\`a}},\ }\href {\doibase
  10.1103/PhysRevA.72.025601} {\bibfield  {journal} {\bibinfo  {journal} {Phys.
  Rev. A}\ }\textbf {\bibinfo {volume} {72}},\ \bibinfo {pages} {025601}
  (\bibinfo {year} {2005})}\BibitemShut {NoStop}%
\bibitem [{\citenamefont {Kinnunen}\ \emph {et~al.}(2006)\citenamefont
  {Kinnunen}, \citenamefont {Jensen},\ and\ \citenamefont
  {T\"orm\"a}}]{PhysRevLett.96.110403}%
  \BibitemOpen
  \bibfield  {author} {\bibinfo {author} {\bibfnamefont {J.}~\bibnamefont
  {Kinnunen}}, \bibinfo {author} {\bibfnamefont {L.~M.}\ \bibnamefont
  {Jensen}}, \ and\ \bibinfo {author} {\bibfnamefont {P.}~\bibnamefont
  {T\"orm\"a}},\ }\href {\doibase 10.1103/PhysRevLett.96.110403} {\bibfield
  {journal} {\bibinfo  {journal} {Phys. Rev. Lett.}\ }\textbf {\bibinfo
  {volume} {96}},\ \bibinfo {pages} {110403} (\bibinfo {year}
  {2006})}\BibitemShut {NoStop}%
\bibitem [{\citenamefont {Mizushima}\ \emph {et~al.}(2007)\citenamefont
  {Mizushima}, \citenamefont {Ichioka},\ and\ \citenamefont
  {Machida}}]{JPSJ.76.104006}%
  \BibitemOpen
  \bibfield  {author} {\bibinfo {author} {\bibfnamefont {T.}~\bibnamefont
  {Mizushima}}, \bibinfo {author} {\bibfnamefont {M.}~\bibnamefont {Ichioka}},
  \ and\ \bibinfo {author} {\bibfnamefont {K.}~\bibnamefont {Machida}},\ }\href
  {\doibase 10.1143/JPSJ.76.104006} {\bibfield  {journal} {\bibinfo  {journal}
  {J. Phys. Soc. Jpn.}\ }\textbf {\bibinfo {volume} {76}},\ \bibinfo {pages}
  {104006} (\bibinfo {year} {2007})}\BibitemShut {NoStop}%
\bibitem [{\citenamefont {Machida}\ \emph {et~al.}(2006)\citenamefont
  {Machida}, \citenamefont {Mizushima},\ and\ \citenamefont
  {Ichioka}}]{PhysRevLett.97.120407}%
  \BibitemOpen
  \bibfield  {author} {\bibinfo {author} {\bibfnamefont {K.}~\bibnamefont
  {Machida}}, \bibinfo {author} {\bibfnamefont {T.}~\bibnamefont {Mizushima}},
  \ and\ \bibinfo {author} {\bibfnamefont {M.}~\bibnamefont {Ichioka}},\ }\href
  {\doibase 10.1103/PhysRevLett.97.120407} {\bibfield  {journal} {\bibinfo
  {journal} {Phys. Rev. Lett.}\ }\textbf {\bibinfo {volume} {97}},\ \bibinfo
  {pages} {120407} (\bibinfo {year} {2006})}\BibitemShut {NoStop}%
\bibitem [{\citenamefont {Liu}\ \emph {et~al.}(2007{\natexlab{a}})\citenamefont
  {Liu}, \citenamefont {Hu},\ and\ \citenamefont
  {Drummond}}]{PhysRevA.75.023614}%
  \BibitemOpen
  \bibfield  {author} {\bibinfo {author} {\bibfnamefont {X.-J.}\ \bibnamefont
  {Liu}}, \bibinfo {author} {\bibfnamefont {H.}~\bibnamefont {Hu}}, \ and\
  \bibinfo {author} {\bibfnamefont {P.~D.}\ \bibnamefont {Drummond}},\ }\href
  {\doibase 10.1103/PhysRevA.75.023614} {\bibfield  {journal} {\bibinfo
  {journal} {Phys. Rev. A}\ }\textbf {\bibinfo {volume} {75}},\ \bibinfo
  {pages} {023614} (\bibinfo {year} {2007}{\natexlab{a}})}\BibitemShut
  {NoStop}%
\bibitem [{\citenamefont {{Tezuka}}\ \emph {et~al.}(2008)\citenamefont
  {{Tezuka}}, \citenamefont {{Yanase}},\ and\ \citenamefont
  {{Ueda}}}]{Tezuka-Yanase}%
  \BibitemOpen
  \bibfield  {author} {\bibinfo {author} {\bibfnamefont {M.}~\bibnamefont
  {{Tezuka}}}, \bibinfo {author} {\bibfnamefont {Y.}~\bibnamefont {{Yanase}}},
  \ and\ \bibinfo {author} {\bibfnamefont {M.}~\bibnamefont {{Ueda}}},\
  }\href@noop {} {\bibfield  {journal} {\bibinfo  {journal} {e-print}\ }
  (\bibinfo {year} {2008})},\ \Eprint {http://arxiv.org/abs/arXiv:0811.1650}
  {arXiv:0811.1650} \BibitemShut {NoStop}%
\bibitem [{\citenamefont {Yanase}(2009)}]{PhysRevB.80.220510}%
  \BibitemOpen
  \bibfield  {author} {\bibinfo {author} {\bibfnamefont {Y.}~\bibnamefont
  {Yanase}},\ }\href {\doibase 10.1103/PhysRevB.80.220510} {\bibfield
  {journal} {\bibinfo  {journal} {Phys. Rev. B}\ }\textbf {\bibinfo {volume}
  {80}},\ \bibinfo {pages} {220510} (\bibinfo {year} {2009})}\BibitemShut
  {NoStop}%
\bibitem [{\citenamefont {Kashimura}\ \emph {et~al.}(2011)\citenamefont
  {Kashimura}, \citenamefont {Tsuchiya},\ and\ \citenamefont
  {Ohashi}}]{PhysRevA.84.013609}%
  \BibitemOpen
  \bibfield  {author} {\bibinfo {author} {\bibfnamefont {T.}~\bibnamefont
  {Kashimura}}, \bibinfo {author} {\bibfnamefont {S.}~\bibnamefont {Tsuchiya}},
  \ and\ \bibinfo {author} {\bibfnamefont {Y.}~\bibnamefont {Ohashi}},\ }\href
  {\doibase 10.1103/PhysRevA.84.013609} {\bibfield  {journal} {\bibinfo
  {journal} {Phys. Rev. A}\ }\textbf {\bibinfo {volume} {84}},\ \bibinfo
  {pages} {013609} (\bibinfo {year} {2011})}\BibitemShut {NoStop}%
\bibitem [{\citenamefont {Nikoli\ifmmode~\acute{c}\else
  \'{c}\fi{}}(2010)}]{PhysRevA.81.023601}%
  \BibitemOpen
  \bibfield  {author} {\bibinfo {author} {\bibfnamefont {P.}~\bibnamefont
  {Nikoli\ifmmode~\acute{c}\else \'{c}\fi{}}},\ }\href {\doibase
  10.1103/PhysRevA.81.023601} {\bibfield  {journal} {\bibinfo  {journal} {Phys.
  Rev. A}\ }\textbf {\bibinfo {volume} {81}},\ \bibinfo {pages} {023601}
  (\bibinfo {year} {2010})}\BibitemShut {NoStop}%
\bibitem [{\citenamefont {Kuli\ifmmode~\acute{c}\else \'{c}\fi{}}\ \emph
  {et~al.}(2009)\citenamefont {Kuli\ifmmode~\acute{c}\else \'{c}\fi{}},
  \citenamefont {Sedrakian},\ and\ \citenamefont
  {Rischke}}]{PhysRevA.80.043610}%
  \BibitemOpen
  \bibfield  {author} {\bibinfo {author} {\bibfnamefont {M.~L.}\ \bibnamefont
  {Kuli\ifmmode~\acute{c}\else \'{c}\fi{}}}, \bibinfo {author} {\bibfnamefont
  {A.}~\bibnamefont {Sedrakian}}, \ and\ \bibinfo {author} {\bibfnamefont
  {D.~H.}\ \bibnamefont {Rischke}},\ }\href {\doibase
  10.1103/PhysRevA.80.043610} {\bibfield  {journal} {\bibinfo  {journal} {Phys.
  Rev. A}\ }\textbf {\bibinfo {volume} {80}},\ \bibinfo {pages} {043610}
  (\bibinfo {year} {2009})}\BibitemShut {NoStop}%
\bibitem [{\citenamefont {Samokhvalov}\ \emph {et~al.}(2010)\citenamefont
  {Samokhvalov}, \citenamefont {Mel'nikov},\ and\ \citenamefont
  {Buzdin}}]{PhysRevB.82.174514}%
  \BibitemOpen
  \bibfield  {author} {\bibinfo {author} {\bibfnamefont {A.~V.}\ \bibnamefont
  {Samokhvalov}}, \bibinfo {author} {\bibfnamefont {A.~S.}\ \bibnamefont
  {Mel'nikov}}, \ and\ \bibinfo {author} {\bibfnamefont {A.~I.}\ \bibnamefont
  {Buzdin}},\ }\href {\doibase 10.1103/PhysRevB.82.174514} {\bibfield
  {journal} {\bibinfo  {journal} {Phys. Rev. B}\ }\textbf {\bibinfo {volume}
  {82}},\ \bibinfo {pages} {174514} (\bibinfo {year} {2010})}\BibitemShut
  {NoStop}%
\bibitem [{\citenamefont {Samokhvalov}\ \emph {et~al.}(2007)\citenamefont
  {Samokhvalov}, \citenamefont {Mel'nikov},\ and\ \citenamefont
  {Buzdin}}]{PhysRevB.76.184519}%
  \BibitemOpen
  \bibfield  {author} {\bibinfo {author} {\bibfnamefont {A.~V.}\ \bibnamefont
  {Samokhvalov}}, \bibinfo {author} {\bibfnamefont {A.~S.}\ \bibnamefont
  {Mel'nikov}}, \ and\ \bibinfo {author} {\bibfnamefont {A.~I.}\ \bibnamefont
  {Buzdin}},\ }\href {\doibase 10.1103/PhysRevB.76.184519} {\bibfield
  {journal} {\bibinfo  {journal} {Phys. Rev. B}\ }\textbf {\bibinfo {volume}
  {76}},\ \bibinfo {pages} {184519} (\bibinfo {year} {2007})}\BibitemShut
  {NoStop}%
\bibitem [{\citenamefont {Zyuzin}\ and\ \citenamefont
  {Zyuzin}(2009)}]{PhysRevB.79.174514}%
  \BibitemOpen
  \bibfield  {author} {\bibinfo {author} {\bibfnamefont {A.~A.}\ \bibnamefont
  {Zyuzin}}\ and\ \bibinfo {author} {\bibfnamefont {A.~Y.}\ \bibnamefont
  {Zyuzin}},\ }\href {\doibase 10.1103/PhysRevB.79.174514} {\bibfield
  {journal} {\bibinfo  {journal} {Phys. Rev. B}\ }\textbf {\bibinfo {volume}
  {79}},\ \bibinfo {pages} {174514} (\bibinfo {year} {2009})}\BibitemShut
  {NoStop}%
\bibitem [{\citenamefont {Quan}\ and\ \citenamefont
  {Zhu}(2010)}]{PhysRevB.81.014518}%
  \BibitemOpen
  \bibfield  {author} {\bibinfo {author} {\bibfnamefont {H.~T.}\ \bibnamefont
  {Quan}}\ and\ \bibinfo {author} {\bibfnamefont {J.-X.}\ \bibnamefont {Zhu}},\
  }\href {\doibase 10.1103/PhysRevB.81.014518} {\bibfield  {journal} {\bibinfo
  {journal} {Phys. Rev. B}\ }\textbf {\bibinfo {volume} {81}},\ \bibinfo
  {pages} {014518} (\bibinfo {year} {2010})}\BibitemShut {NoStop}%
\bibitem [{\citenamefont {Ryu}\ \emph {et~al.}(2007)\citenamefont {Ryu},
  \citenamefont {Andersen}, \citenamefont {Clad\'{e}}, \citenamefont
  {Natarajan}, \citenamefont {Helmerson},\ and\ \citenamefont
  {Phillips}}]{ryu:260401}%
  \BibitemOpen
  \bibfield  {author} {\bibinfo {author} {\bibfnamefont {C.}~\bibnamefont
  {Ryu}}, \bibinfo {author} {\bibfnamefont {M.~F.}\ \bibnamefont {Andersen}},
  \bibinfo {author} {\bibfnamefont {P.}~\bibnamefont {Clad\'{e}}}, \bibinfo
  {author} {\bibfnamefont {V.}~\bibnamefont {Natarajan}}, \bibinfo {author}
  {\bibfnamefont {K.}~\bibnamefont {Helmerson}}, \ and\ \bibinfo {author}
  {\bibfnamefont {W.~D.}\ \bibnamefont {Phillips}},\ }\href {\doibase
  10.1103/PhysRevLett.99.260401} {\bibfield  {journal} {\bibinfo  {journal}
  {Phys. Rev. Lett.}\ }\textbf {\bibinfo {volume} {99}},\ \bibinfo {eid}
  {260401} (\bibinfo {year} {2007})}\BibitemShut {NoStop}%
\bibitem [{\citenamefont {Sherlock}\ \emph {et~al.}(2011)\citenamefont
  {Sherlock}, \citenamefont {Gildemeister}, \citenamefont {Owen}, \citenamefont
  {Nugent},\ and\ \citenamefont {Foot}}]{Sherlock}%
  \BibitemOpen
  \bibfield  {author} {\bibinfo {author} {\bibfnamefont {B.~E.}\ \bibnamefont
  {Sherlock}}, \bibinfo {author} {\bibfnamefont {M.}~\bibnamefont
  {Gildemeister}}, \bibinfo {author} {\bibfnamefont {E.}~\bibnamefont {Owen}},
  \bibinfo {author} {\bibfnamefont {E.}~\bibnamefont {Nugent}}, \ and\ \bibinfo
  {author} {\bibfnamefont {C.~J.}\ \bibnamefont {Foot}},\ }\href {\doibase
  10.1103/PhysRevA.83.043408} {\bibfield  {journal} {\bibinfo  {journal} {Phys.
  Rev. A}\ }\textbf {\bibinfo {volume} {83}},\ \bibinfo {pages} {043408}
  (\bibinfo {year} {2011})}\BibitemShut {NoStop}%
\bibitem [{\citenamefont {Yang}(2001)}]{PhysRevB.63.140511}%
  \BibitemOpen
  \bibfield  {author} {\bibinfo {author} {\bibfnamefont {K.}~\bibnamefont
  {Yang}},\ }\href {\doibase 10.1103/PhysRevB.63.140511} {\bibfield  {journal}
  {\bibinfo  {journal} {Phys. Rev. B}\ }\textbf {\bibinfo {volume} {63}},\
  \bibinfo {pages} {140511} (\bibinfo {year} {2001})}\BibitemShut {NoStop}%
\bibitem [{\citenamefont {Orso}(2007)}]{orso:070402}%
  \BibitemOpen
  \bibfield  {author} {\bibinfo {author} {\bibfnamefont {G.}~\bibnamefont
  {Orso}},\ }\href {\doibase 10.1103/PhysRevLett.98.070402} {\bibfield
  {journal} {\bibinfo  {journal} {Phys. Rev. Lett.}\ }\textbf {\bibinfo
  {volume} {98}},\ \bibinfo {eid} {070402} (\bibinfo {year}
  {2007})}\BibitemShut {NoStop}%
\bibitem [{\citenamefont {Tezuka}\ and\ \citenamefont
  {Ueda}(2008)}]{tezuka:110403}%
  \BibitemOpen
  \bibfield  {author} {\bibinfo {author} {\bibfnamefont {M.}~\bibnamefont
  {Tezuka}}\ and\ \bibinfo {author} {\bibfnamefont {M.}~\bibnamefont {Ueda}},\
  }\href {\doibase 10.1103/PhysRevLett.100.110403} {\bibfield  {journal}
  {\bibinfo  {journal} {Phys. Rev. Lett.}\ }\textbf {\bibinfo {volume} {100}},\
  \bibinfo {eid} {110403} (\bibinfo {year} {2008})}\BibitemShut {NoStop}%
\bibitem [{\citenamefont {Hu}\ \emph {et~al.}(2007)\citenamefont {Hu},
  \citenamefont {Liu},\ and\ \citenamefont {Drummond}}]{hu:070403}%
  \BibitemOpen
  \bibfield  {author} {\bibinfo {author} {\bibfnamefont {H.}~\bibnamefont
  {Hu}}, \bibinfo {author} {\bibfnamefont {X.-J.}\ \bibnamefont {Liu}}, \ and\
  \bibinfo {author} {\bibfnamefont {P.~D.}\ \bibnamefont {Drummond}},\ }\href
  {\doibase 10.1103/PhysRevLett.98.070403} {\bibfield  {journal} {\bibinfo
  {journal} {Phys. Rev. Lett.}\ }\textbf {\bibinfo {volume} {98}},\ \bibinfo
  {eid} {070403} (\bibinfo {year} {2007})}\BibitemShut {NoStop}%
\bibitem [{\citenamefont {Liu}\ \emph {et~al.}(2007{\natexlab{b}})\citenamefont
  {Liu}, \citenamefont {Hu},\ and\ \citenamefont {Drummond}}]{liu:043605}%
  \BibitemOpen
  \bibfield  {author} {\bibinfo {author} {\bibfnamefont {X.-J.}\ \bibnamefont
  {Liu}}, \bibinfo {author} {\bibfnamefont {H.}~\bibnamefont {Hu}}, \ and\
  \bibinfo {author} {\bibfnamefont {P.~D.}\ \bibnamefont {Drummond}},\ }\href
  {\doibase 10.1103/PhysRevA.76.043605} {\bibfield  {journal} {\bibinfo
  {journal} {Phys. Rev. A}\ }\textbf {\bibinfo {volume} {76}},\ \bibinfo
  {pages} {043605} (\bibinfo {year} {2007}{\natexlab{b}})}\BibitemShut
  {NoStop}%
\bibitem [{\citenamefont {Feiguin}\ and\ \citenamefont
  {Heidrich-Meisner}(2007)}]{feiguin:220508}%
  \BibitemOpen
  \bibfield  {author} {\bibinfo {author} {\bibfnamefont {A.~E.}\ \bibnamefont
  {Feiguin}}\ and\ \bibinfo {author} {\bibfnamefont {F.}~\bibnamefont
  {Heidrich-Meisner}},\ }\href {\doibase 10.1103/PhysRevB.76.220508} {\bibfield
   {journal} {\bibinfo  {journal} {Phys. Rev. B}\ }\textbf {\bibinfo {volume}
  {76}},\ \bibinfo {eid} {220508} (\bibinfo {year} {2007})}\BibitemShut
  {NoStop}%
\bibitem [{\citenamefont {Feiguin}\ and\ \citenamefont
  {Heidrich-Meisner}(2009)}]{feiguin:076403}%
  \BibitemOpen
  \bibfield  {author} {\bibinfo {author} {\bibfnamefont {A.~E.}\ \bibnamefont
  {Feiguin}}\ and\ \bibinfo {author} {\bibfnamefont {F.}~\bibnamefont
  {Heidrich-Meisner}},\ }\href {\doibase 10.1103/PhysRevLett.102.076403}
  {\bibfield  {journal} {\bibinfo  {journal} {Phys. Rev. Lett.}\ }\textbf
  {\bibinfo {volume} {102}},\ \bibinfo {pages} {076403} (\bibinfo {year}
  {2009})}\BibitemShut {NoStop}%
\bibitem [{\citenamefont {Rizzi}\ \emph {et~al.}(2008)\citenamefont {Rizzi},
  \citenamefont {Polini}, \citenamefont {Cazalilla}, \citenamefont {Bakhtiari},
  \citenamefont {Tosi},\ and\ \citenamefont {Fazio}}]{rizzi:245105}%
  \BibitemOpen
  \bibfield  {author} {\bibinfo {author} {\bibfnamefont {M.}~\bibnamefont
  {Rizzi}}, \bibinfo {author} {\bibfnamefont {M.}~\bibnamefont {Polini}},
  \bibinfo {author} {\bibfnamefont {M.~A.}\ \bibnamefont {Cazalilla}}, \bibinfo
  {author} {\bibfnamefont {M.~R.}\ \bibnamefont {Bakhtiari}}, \bibinfo {author}
  {\bibfnamefont {M.~P.}\ \bibnamefont {Tosi}}, \ and\ \bibinfo {author}
  {\bibfnamefont {R.}~\bibnamefont {Fazio}},\ }\href {\doibase
  10.1103/PhysRevB.77.245105} {\bibfield  {journal} {\bibinfo  {journal} {Phys.
  Rev. B}\ }\textbf {\bibinfo {volume} {77}},\ \bibinfo {eid} {245105}
  (\bibinfo {year} {2008})}\BibitemShut {NoStop}%
\bibitem [{\citenamefont {Parish}\ \emph {et~al.}(2007)\citenamefont {Parish},
  \citenamefont {Baur}, \citenamefont {Mueller},\ and\ \citenamefont
  {Huse}}]{parish:250403}%
  \BibitemOpen
  \bibfield  {author} {\bibinfo {author} {\bibfnamefont {M.~M.}\ \bibnamefont
  {Parish}}, \bibinfo {author} {\bibfnamefont {S.~K.}\ \bibnamefont {Baur}},
  \bibinfo {author} {\bibfnamefont {E.~J.}\ \bibnamefont {Mueller}}, \ and\
  \bibinfo {author} {\bibfnamefont {D.~A.}\ \bibnamefont {Huse}},\ }\href
  {\doibase 10.1103/PhysRevLett.99.250403} {\bibfield  {journal} {\bibinfo
  {journal} {Phys. Rev. Lett.}\ }\textbf {\bibinfo {volume} {99}},\ \bibinfo
  {eid} {250403} (\bibinfo {year} {2007})}\BibitemShut {NoStop}%
\bibitem [{\citenamefont {Zhao}\ and\ \citenamefont {Liu}(2008)}]{zhao:063605}%
  \BibitemOpen
  \bibfield  {author} {\bibinfo {author} {\bibfnamefont {E.}~\bibnamefont
  {Zhao}}\ and\ \bibinfo {author} {\bibfnamefont {W.~V.}\ \bibnamefont {Liu}},\
  }\href {\doibase 10.1103/PhysRevA.78.063605} {\bibfield  {journal} {\bibinfo
  {journal} {Phys. Rev. A}\ }\textbf {\bibinfo {volume} {78}},\ \bibinfo {eid}
  {063605} (\bibinfo {year} {2008})}\BibitemShut {NoStop}%
\bibitem [{\citenamefont {Lutchyn}\ \emph {et~al.}(2011)\citenamefont
  {Lutchyn}, \citenamefont {Dzero},\ and\ \citenamefont
  {Yakovenko}}]{PhysRevA.84.033609}%
  \BibitemOpen
  \bibfield  {author} {\bibinfo {author} {\bibfnamefont {R.~M.}\ \bibnamefont
  {Lutchyn}}, \bibinfo {author} {\bibfnamefont {M.}~\bibnamefont {Dzero}}, \
  and\ \bibinfo {author} {\bibfnamefont {V.~M.}\ \bibnamefont {Yakovenko}},\
  }\href {\doibase 10.1103/PhysRevA.84.033609} {\bibfield  {journal} {\bibinfo
  {journal} {Phys. Rev. A}\ }\textbf {\bibinfo {volume} {84}},\ \bibinfo
  {pages} {033609} (\bibinfo {year} {2011})}\BibitemShut {NoStop}%
\bibitem [{\citenamefont {Chen}\ \emph {et~al.}(2005)\citenamefont {Chen},
  \citenamefont {Stajic}, \citenamefont {Tan},\ and\ \citenamefont
  {Levin}}]{Chen20051}%
  \BibitemOpen
  \bibfield  {author} {\bibinfo {author} {\bibfnamefont {Q.}~\bibnamefont
  {Chen}}, \bibinfo {author} {\bibfnamefont {J.}~\bibnamefont {Stajic}},
  \bibinfo {author} {\bibfnamefont {S.}~\bibnamefont {Tan}}, \ and\ \bibinfo
  {author} {\bibfnamefont {K.}~\bibnamefont {Levin}},\ }\href {\doibase DOI:
  10.1016/j.physrep.2005.02.005} {\bibfield  {journal} {\bibinfo  {journal}
  {Phys. Rep.}\ }\textbf {\bibinfo {volume} {412}},\ \bibinfo {pages} {1 }
  (\bibinfo {year} {2005})}\BibitemShut {NoStop}%
\bibitem [{\citenamefont {Heidrich-Meisner}\ \emph {et~al.}(2010)\citenamefont
  {Heidrich-Meisner}, \citenamefont {Feiguin}, \citenamefont {Schollw\"ock},\
  and\ \citenamefont {Zwerger}}]{meisner:023629}%
  \BibitemOpen
  \bibfield  {author} {\bibinfo {author} {\bibfnamefont {F.}~\bibnamefont
  {Heidrich-Meisner}}, \bibinfo {author} {\bibfnamefont {A.~E.}\ \bibnamefont
  {Feiguin}}, \bibinfo {author} {\bibfnamefont {U.}~\bibnamefont
  {Schollw\"ock}}, \ and\ \bibinfo {author} {\bibfnamefont {W.}~\bibnamefont
  {Zwerger}},\ }\href {\doibase 10.1103/PhysRevA.81.023629} {\bibfield
  {journal} {\bibinfo  {journal} {Phys. Rev. A}\ }\textbf {\bibinfo {volume}
  {81}},\ \bibinfo {pages} {023629} (\bibinfo {year} {2010})}\BibitemShut
  {NoStop}%
\bibitem [{\citenamefont {Bulgac}\ and\ \citenamefont
  {Forbes}(2008)}]{bulgac:215301}%
  \BibitemOpen
  \bibfield  {author} {\bibinfo {author} {\bibfnamefont {A.}~\bibnamefont
  {Bulgac}}\ and\ \bibinfo {author} {\bibfnamefont {M.~M.}\ \bibnamefont
  {Forbes}},\ }\href {\doibase 10.1103/PhysRevLett.101.215301} {\bibfield
  {journal} {\bibinfo  {journal} {Phys. Rev. Lett.}\ }\textbf {\bibinfo
  {volume} {101}},\ \bibinfo {pages} {215301} (\bibinfo {year}
  {2008})}\BibitemShut {NoStop}%
\bibitem [{\citenamefont {Machida}\ and\ \citenamefont
  {Nakanishi}(1984)}]{machida1984}%
  \BibitemOpen
  \bibfield  {author} {\bibinfo {author} {\bibfnamefont {K.}~\bibnamefont
  {Machida}}\ and\ \bibinfo {author} {\bibfnamefont {H.}~\bibnamefont
  {Nakanishi}},\ }\href {\doibase 10.1103/PhysRevB.30.122} {\bibfield
  {journal} {\bibinfo  {journal} {Phys. Rev. B}\ }\textbf {\bibinfo {volume}
  {30}},\ \bibinfo {pages} {122} (\bibinfo {year} {1984})}\BibitemShut
  {NoStop}%
\bibitem [{\citenamefont {Yoshii}\ \emph {et~al.}(2011)\citenamefont {Yoshii},
  \citenamefont {Tsuchiya}, \citenamefont {Marmorini},\ and\ \citenamefont
  {Nitta}}]{PhysRevB.84.024503}%
  \BibitemOpen
  \bibfield  {author} {\bibinfo {author} {\bibfnamefont {R.}~\bibnamefont
  {Yoshii}}, \bibinfo {author} {\bibfnamefont {S.}~\bibnamefont {Tsuchiya}},
  \bibinfo {author} {\bibfnamefont {G.}~\bibnamefont {Marmorini}}, \ and\
  \bibinfo {author} {\bibfnamefont {M.}~\bibnamefont {Nitta}},\ }\href
  {\doibase 10.1103/PhysRevB.84.024503} {\bibfield  {journal} {\bibinfo
  {journal} {Phys. Rev. B}\ }\textbf {\bibinfo {volume} {84}},\ \bibinfo
  {pages} {024503} (\bibinfo {year} {2011})}\BibitemShut {NoStop}%
\bibitem [{\citenamefont {Vorontsov}\ and\ \citenamefont
  {Graf}(2006)}]{PhysRevB.74.172504}%
  \BibitemOpen
  \bibfield  {author} {\bibinfo {author} {\bibfnamefont {A.~B.}\ \bibnamefont
  {Vorontsov}}\ and\ \bibinfo {author} {\bibfnamefont {M.~J.}\ \bibnamefont
  {Graf}},\ }\href {\doibase 10.1103/PhysRevB.74.172504} {\bibfield  {journal}
  {\bibinfo  {journal} {Phys. Rev. B}\ }\textbf {\bibinfo {volume} {74}},\
  \bibinfo {pages} {172504} (\bibinfo {year} {2006})}\BibitemShut {NoStop}%
\bibitem [{\citenamefont {Ivanov}(2001)}]{ivanov2001}%
  \BibitemOpen
  \bibfield  {author} {\bibinfo {author} {\bibfnamefont {D.~A.}\ \bibnamefont
  {Ivanov}},\ }\href {\doibase 10.1103/PhysRevLett.86.268} {\bibfield
  {journal} {\bibinfo  {journal} {Phys. Rev. Lett.}\ }\textbf {\bibinfo
  {volume} {86}},\ \bibinfo {pages} {268} (\bibinfo {year} {2001})}\BibitemShut
  {NoStop}%
\bibitem [{\citenamefont {Sudb\o{}}\ \emph {et~al.}(1993)\citenamefont
  {Sudb\o{}}, \citenamefont {Varma}, \citenamefont {Giamarchi}, \citenamefont
  {Stechel},\ and\ \citenamefont {Scalettar}}]{PhysRevLett.70.978}%
  \BibitemOpen
  \bibfield  {author} {\bibinfo {author} {\bibfnamefont {A.}~\bibnamefont
  {Sudb\o{}}}, \bibinfo {author} {\bibfnamefont {C.~M.}\ \bibnamefont {Varma}},
  \bibinfo {author} {\bibfnamefont {T.}~\bibnamefont {Giamarchi}}, \bibinfo
  {author} {\bibfnamefont {E.~B.}\ \bibnamefont {Stechel}}, \ and\ \bibinfo
  {author} {\bibfnamefont {R.~T.}\ \bibnamefont {Scalettar}},\ }\href {\doibase
  10.1103/PhysRevLett.70.978} {\bibfield  {journal} {\bibinfo  {journal} {Phys.
  Rev. Lett.}\ }\textbf {\bibinfo {volume} {70}},\ \bibinfo {pages} {978}
  (\bibinfo {year} {1993})}\BibitemShut {NoStop}%
\bibitem [{\citenamefont {Chen}\ \emph {et~al.}(2009)\citenamefont {Chen},
  \citenamefont {Wang}, \citenamefont {Zhang},\ and\ \citenamefont
  {Ting}}]{PhysRevB.79.054512}%
  \BibitemOpen
  \bibfield  {author} {\bibinfo {author} {\bibfnamefont {Y.}~\bibnamefont
  {Chen}}, \bibinfo {author} {\bibfnamefont {Z.~D.}\ \bibnamefont {Wang}},
  \bibinfo {author} {\bibfnamefont {F.~C.}\ \bibnamefont {Zhang}}, \ and\
  \bibinfo {author} {\bibfnamefont {C.~S.}\ \bibnamefont {Ting}},\ }\href
  {\doibase 10.1103/PhysRevB.79.054512} {\bibfield  {journal} {\bibinfo
  {journal} {Phys. Rev. B}\ }\textbf {\bibinfo {volume} {79}},\ \bibinfo
  {pages} {054512} (\bibinfo {year} {2009})}\BibitemShut {NoStop}%
\bibitem [{\citenamefont {Agterberg}\ and\ \citenamefont
  {Kaur}(2007)}]{Agterberg_helical}%
  \BibitemOpen
  \bibfield  {author} {\bibinfo {author} {\bibfnamefont {D.~F.}\ \bibnamefont
  {Agterberg}}\ and\ \bibinfo {author} {\bibfnamefont {R.~P.}\ \bibnamefont
  {Kaur}},\ }\href {\doibase 10.1103/PhysRevB.75.064511} {\bibfield  {journal}
  {\bibinfo  {journal} {Phys. Rev. B}\ }\textbf {\bibinfo {volume} {75}},\
  \bibinfo {pages} {064511} (\bibinfo {year} {2007})}\BibitemShut {NoStop}%
\end{thebibliography}%
\end{document}